\documentclass[prb,aps,amssymb,amsmath,twocolumn,showpacs,floatfix]{revtex4}
\usepackage[dvips]{graphicx}
\usepackage{mathrsfs}

\def\be{\begin{equation}}
\def\ee{\end{equation}}
\def\bea{\begin{eqnarray}}
\def\eea{\end{eqnarray}}
\def\ve{\varepsilon}

\begin{document}

\title{Quantum interference and spin-charge separation \\
in a disordered Luttinger liquid}
\author{A.G.~Yashenkin$^{1,2}$}
\author{I.V.~Gornyi$^{1,3}$}
\author{A.D.~Mirlin$^{1,4,2}$}
\author{D.G.~Polyakov$^{1}$}
\affiliation{$^{1}$Institut f\"ur Nanotechnologie,
Forschungszentrum Karlsruhe, 76021 Karlsruhe, Germany \\
$^{2}$Theory Division, Petersburg Nuclear Physics Institute, Gatchina,
188300 St.~Petersburg,  Russia \\
$^{3}$A.F.~Ioffe Physico-Technical Institute, 194021 St.~Petersburg, Russia\\
$^{4}$Institut f\"ur Theorie der kondensierten Materie, Universit\"at
Karlsruhe, 76128 Karlsruhe, Germany}

\date{\today}

\begin{abstract}
We study the influence of spin on the quantum interference of interacting
electrons in a single-channel disordered quantum wire within the framework of
the Luttinger liquid (LL) model. The nature of the electron interference in a
spinful LL is particularly nontrivial because the elementary bosonic
excitations that carry charge and spin propagate with different velocities. We
extend the functional bosonization approach to treat the fermionic and bosonic
degrees of freedom in a disordered spinful LL on an equal footing. We analyze
the effect of spin-charge separation at finite temperature both on the
spectral properties of single-particle fermionic excitations and on the
conductivity of a disordered quantum wire. We demonstrate that the notion of
weak localization, related to the interference of multiple-scattered electron
waves and their decoherence due to electron-electron scattering, remains
applicable to the spin-charge separated system. The relevant dephasing length,
governed by the interplay of electron-electron interaction and spin-charge
separation, is found to be parametrically shorter than in a spinless
LL. We calculate both the quantum (weak localization) and classical
(memory effect) corrections to the conductivity of a disordered spinful
LL. The classical correction is shown to dominate in the limit of high
temperature.
\end{abstract}


\maketitle

\section{Introduction}
\label{Intro}

Interacting electrons in one dimension (1D) are a paradigmatic example of the
strongly correlated fermionic systems. Electron-electron (e-e) interactions
drive the 1D system into a non-Fermi liquid state known as the Luttinger
liquid (LL) (for review see, e.g.,
Refs.~\onlinecite{sol79,voit94,schulz95,gog98,schulz00,maslov,giam04}). In
recent years, progress in nanofabrication technologies has made it possible to
manufacture a variety of single- and few-channel quantum wires connected to
the electric leads and to perform systematic transport measurements on the
very narrow wires. The latter include single--wall carbon nanotubes
\cite{nano1,nano2,nano3,nano4,nano5,nano6,nano7},
semiconductor-based\cite{semicond1,semicond2,Jompol,Hew08} and
metallic\cite{metallic} quantum wires, polymer nanofibers\cite{polymer}, as
well as quantum Hall edge states.\cite{QHedge,grayson05} The LL nature of
these strongly correlated quantum wires has been supported by a wealth of
experimental
findings.\cite{nano1,nano2,nano3,nano4,nano5,nano6,nano7,semicond1,semicond2,
metallic,polymer,QHedge,grayson05} Another
class of strongly-correlated quantum wires that has recently attracted a lot of
interest
is ultracold atomic gases confined to 1D geometry,
for review see Ref.~\onlinecite{bloch08}.

Mesoscopic physics of strongly correlated electrons is one of the most
important and promising directions of current research on 1D electron
systems. Recent transport measurements on carbon nanotubes\cite{morpurgo}
reported both sample-dependent conductance fluctuations and strong
magnetoconductivity, in qualitative similarity to the mesoscopic phenomena in
higher-dimensional disordered electron systems. In Ref.~\onlinecite{morpurgo},
the sample size $\sim 1\:\mu$m was of the order of the mean free path limited
by impurity scattering. Electron transport through the nanotubes displayed,
therefore, features characteristic of the crossover from ballistic conduction
to a disorder-dominated regime. On the other hand, in the past few years
techniques to grow nanotubes of size up to the millimeter
scale\cite{li04,purewal07}---much larger than the typical value of the
disorder-induced mean free path---have been developed. First transport
measurements\cite{li04,purewal07} on the ultralong nanotubes provided evidence
for disorder-induced {\it diffusive} motion of electrons in a wide range of
temperature. Altogether, these advances have paved the way for systematic
experimental study of interference-induced localization phenomena in 1D.

The theory of weak localization (WL) in a disordered LL of {\it
spinless} electrons was developed in Refs.~\onlinecite{GMPlett} and
\onlinecite{GMP} (for recent advances in the 
ballistic $\sigma$-model framework, see also
Ref.~\onlinecite{micklitz08}). In the limit of weak interaction
between electrons, the phase breaking length $l_\phi$ that sets up the
infrared cutoff of WL was shown to obey $l_{\phi} \sim\alpha^{-1}
(l_Tl)^{1/2}$, where $\alpha\ll 1$ is the dimensionless interaction
constant, $l_T\sim v/T$ the thermal length, $v$ the Fermi velocity,
$T$ the temperature, $l$ the elastic mean free path. At sufficiently
high temperatures, when $l_{\phi} \ll l$, the system is in the WL
regime.\cite{SL} The WL correction $\Delta\sigma_{\rm WL}$ to the Drude
conductivity $\sigma_{\rm D}$ behaves then as $\Delta\sigma_{\rm WL} / \sigma_{\rm D}
\sim \, -(l_{\phi}/l) ^2 \ln (l/ l_{\phi}) \sim - \alpha^{-2}(l_T/l) \ln
(\alpha^2 l /l_T)$. Note that the WL dephasing length of spinless
electrons originates from the interplay of interaction and disorder:
both ingredients are necessary to establish a nonzero dephasing
rate. As a result, $l_\phi$ in the WL regime is much longer than the
total decay length of fermionic excitations
$l_{ee}\sim\alpha^{-2}l_T$. The latter is the dephasing length
relevant to the damping of Aharonov-Bohm
oscillations\cite{GMPlett,GMP,lehur02,lehur06} and to the smearing of
a zero-bias anomaly in the tunnelling density of states.\cite{gutman08}

In a {\it spinful} LL, the most prominent spin-related manifestation of
non-Fermi liquid physics is a phenomenon of spin-charge separation (SCS) (see,
e.g.,
Refs.~\onlinecite{sol79,voit94,schulz95,gog98,schulz00,maslov,giam04}). The
essence of the SCS in the LL model is that the spin and charge sectors of the
theory in a bosonic representation are completely decoupled from each other
and characterized by different interaction coupling
constants. Correspondingly, the elementary bosonic excitations carrying spin
and charge propagate with different velocities, independently of each
other. Experimentally, the effect of the SCS on the spectral properties of a
LL (as measured in electron tunnelling experiments) has been studied in
Refs.~\onlinecite{semicond1,Jompol}. In the last few years, much attention has been
given to the so-called ``spin-incoherent regime"~\cite{fiete07}
in 1D systems with strongly different spin and charge 
velocities.\cite{matveev04,fiete04,cheianov04} For recent
transport measurements in ballistic quantum wires which show signatures of the
spin-incoherent behavior, see Ref.~\onlinecite{Hew08} and references therein.

While the difference in the spectral properties of spin and charge
collective modes is not at all specific to 1D and is also
characteristic of higher-dimensional Fermi liquids,\cite{baym04} the
peculiarity of 1D is that the ``factorization" of the {\it bosonic}
modes modifies the single-particle {\it fermionic} properties in an
essential way.\cite{footnote1} Our purpose is to
investigate how the SCS affects the quantum interference phenomena in a
disordered LL. In particular, we analyze the dynamical properties of fermionic
excitations at finite temperature and employ the results of this analysis to
calculate the WL correction to the conductivity of a spinful LL. The problem
is rather nontrivial conceptually since the spin and charge degrees of freedom
that constitute the electron in a LL acquire different velocities and the very
notion of a specific quasiclassical electron trajectory characterized by a
certain velocity---conventionally invoked in a description of WL---becomes
ambiguous.

We show below that, despite the intricate nature of a single-electron
motion in a spinful LL, the basic notions of WL remain applicable even
when the SCS is incorporated in the calculation. However, the spin
degree of freedom has dramatic consequences for the effects of e-e
scattering. Most importantly, the decay rate of fermionic excitations
is strongly enhanced in the presence of spin: the single-particle
length $l_{ee}\sim\alpha^{-1}l_T$ becomes parametrically shorter than
for spinless (spin-polarized) electrons. Moreover, in contrast to the
spinless case, the WL dephasing length $l_\phi\sim\alpha^{-1}l_T$
becomes of the order of $l_{ee}$. As a result, also the WL dephasing
turns out to be much stronger than without the SCS being included. The
temperature at which spinful electrons get strongly localized
($l_T/l\sim\alpha$) is therefore much lower (for small $\alpha$) than
for spinless electrons. Furthermore, we find that the WL correction to
the conductivity is given by $\Delta\sigma_{\rm WL}/\sigma_D \sim
-(l_{\phi}/l)^2 \sim - \alpha^{-2}(l_T/l)^{2}$, showing a much faster
temperature dependence of $\Delta\sigma_{\rm WL}$ than in the spinless
case.

We also demonstrate that a classical ``memory effect" (ME) in the electron
scattering off disorder contributes to the $T$ dependence of the conductivity.
Moreover, it gives the leading (larger than $\Delta\sigma_{\rm WL}$) correction to the Drude
conductivity in the limit of high $T$. The obtained ME contribution
$\Delta\sigma_{\rm ME}/\sigma_D\sim - l_T/l$ is essentially not related to e-e
interactions and exceeds $\Delta\sigma_{\rm WL}$ only when the latter is
sufficiently suppressed by the interaction-induced dephasing. Specifically,
$|\Delta\sigma_{\rm ME}|\gg|\Delta\sigma_{\rm WL}|$ for $l_T/l\ll\alpha^2$ (i.e., for
$l_\phi/l\ll\alpha$). Otherwise, $|\Delta\sigma_{\rm WL}|\gg|\Delta\sigma_{\rm ME}|$.
Technically, the ME manifests itself in the same set of Feynman diagrams for
the conductivity as the WL, so that we actually treat the two effects---the
essentially classical ME and the essentially quantum WL---on an equal
footing. What distinguishes them from each other is that the main
contributions to $\Delta\sigma_{\rm WL}$ and $\Delta\sigma_{\rm ME}$ come from
scattering on different impurity configurations. The WL correction stems from
scattering on rare, compact three-impurity complexes in which the
characteristic distance between all three impurities is of the order of the
single-particle length $l_{ee}$. The ME correction is associated with impurity
configurations in which two of impurities are located very close to each
other, with a characteristic distance between them of the order of the thermal
length $l_T$.

The remainder of the paper is organized as follows. In Sec.~\ref{II} we
formulate the model of a disordered LL, discuss the basic physics of the SCS
relevant to the e-e scattering, and describe the method of functional
bosonization used in our calculation. Section \ref{III} is devoted to an
analysis of the spectral properties of fermionic excitations in various
representations; in particular, in a ``space-energy representation" employed
for the calculation of the conductivity. In Secs.~\ref{IV} and \ref{V} we
evaluate the WL and ME corrections to the conductivity, respectively, by means
of the functional bosonization method. In Sec.~\ref{VI} we present a
complementary analysis based on the more conventional path-integral
approach. Our results are summarized in Sec.~\ref{VII}.

\section{Model and method}
\label{II}
\setcounter{equation}{0}

We begin by formulating the model of a disordered spinful LL in
Sec.~\ref{IIa}. In Sec.~\ref{IIaa} we present a simple argument which
demonstrates the peculiarity of 1D geometry in that the spin degree of freedom
affects the rate of e-e scattering in the LL in a crucial way. Section
\ref{IIb} is devoted to an overview of the functional bosonization method.

\subsection{Disordered Luttinger liquid}
\label{IIa}

Throughout the paper we consider a single-channel infinite quantum
wire. Linearizing the dispersion relation of electrons about two Fermi points
at the wavevectors $k=\pm k_F$ with the velocity $v$, the Hamiltonian of a
clean LL is written as ($\hbar=1$)
\begin{eqnarray}
H_{\rm LL} &=& \sum_{k\mu\sigma}v(\mu k - k_F)\Psi^\dagger_{\mu\sigma}(k)
\Psi_{\mu\sigma}(k) \nonumber \\ &+& {1\over
2}\sum_{\mu\sigma\sigma'}\int \!dx \,\big( n_{\mu\sigma}\, g_4 \,
n_{\mu\sigma'} + n_{\mu\sigma} \, g_2 \, n_{-\mu,\sigma'}\big)~.
\nonumber \\
\label{1}
\end{eqnarray}
Here $\Psi_{\mu\sigma}(k)$ are the electron operators at the wavevector $k$,
the index $\mu=\pm$ denotes two branches of chiral excitations (right and left
movers), and $\sigma=\uparrow,\downarrow$ stands for two spin projections.
The e-e interaction enters Eq.~(\ref{1}) through the coupling constants $g_4$ and
$g_2$. These describe forward e-e scattering with small momentum transfer
(much smaller than $k_F$) between electrons from the same ($g_4$) or different
($g_2$) chiral branches. We assume that the e-e interaction is short-ranged
and represent the interaction part of the Hamiltonian in terms of the local
in space electron density operators $n_{\mu\sigma}(x)=
\psi^\dagger_{\mu\sigma}(x) \psi_{\mu\sigma}(x)$.  The parameters of the
Hamiltonian with the linearized dispersion should be understood as effective
(phenomenological) couplings of the low-energy theory, which include possible
high-energy renormalization effects, similar to Fermi-liquid theory.

The LL Hamiltonian (\ref{1}) does not contain the term
\be
H_{\rm bs} =
{1\over 2}\sum_{\mu\sigma\sigma'}\int \!dx\,
\psi^\dagger_{\mu\sigma}
\psi_{-\mu,\sigma} \, g_1 \,
\psi^\dagger_{-\mu,\sigma'} \psi_{\mu\sigma'}~,
\label{1'}
\ee
which describes backward e-e scattering with large momentum transfer resulting
in a change of chirality $\mu$.  If one begins with a microscopic model of electrons
interacting via a finite-range externally screened Coulomb potential, the
constants $g_{2,4}$ in Eq.~(\ref{1}) and $g_1$ in Eq.~(\ref{1'}) are related
to the Fourier transforms of this potential at zero and $2 k_F$ momenta,
respectively. The forward scattering dominates provided that the radius of
external screening $d$ (e.g., the distance to a metallic gate) is much larger
than $k_F^{-1}$. We assume that this is the case and neglect $H_{\rm bs}$
throughout the paper below.\cite{remark1} Treating the Coulomb potential
in Eq.~(\ref{1}) as short-ranged is legitimate for scattering processes with
momentum transfer much smaller than $d^{-1}$. This same scale $d^{-1}$ fixes
the ultraviolet momentum cutoff in our formulation of the low-energy theory.

The only source of electron backscattering in our model is thus a static
random potential $U(x)$ due to the presence of impurities. We assume that
fluctuations of $U(x)$ are Gaussian and characterized by the correlation
function $\langle U(x)U(x') \rangle = \delta(x-x') v^2/2l_0$
(``white noise"). Here $l_0$ is the transport elastic mean free path in the absence
of interaction. The disorder is considered to be weak, $k_F l_0\gg 1$. The
disorder-induced backscattering term in the Hamiltonian is given by
\be
H_{\rm imp} = \sum_{\sigma}\ \int\! dx\,
\big( U_b^* \ \psi^\dagger_{+ \sigma} \psi_{- \sigma} +
 U_b\ \psi^\dagger_{- \sigma}\psi_{+ \sigma}\big)~,
\label{2}
\ee
where the backscattering amplitudes $U_b (x)$ are correlated as $\langle
U_b(x) U^*_b(x')\rangle = \langle U(x)U(x') \rangle $ and $\langle U_b(x) U_b
(x') \rangle = 0 $. Forward scattering off impurities can be gauged out in the
calculation of the conductivity\cite{abrikosov78,giam04} and will therefore be
neglected from the very beginning. The total Hamiltonian $H$ that defines our
model of a disordered spinful LL is thus
\be
H=H_{\rm LL}+H_{\rm imp}~.
\ee

Throughout the paper we consider a quantum wire with spin and chiral
channels not separated spatially in the transverse direction, so that the
constants $g_2=g_4\equiv g$ are spin-independent and equal to each other. The
plasmon velocity $u$ for the spinful case then reads
\be
u=v/K_\rho=v(1+2g/\pi v)^{1/2}~,
\label{2.6}
\ee
with $K_\rho$ being the Luttinger constant in the charge sector,
whereas the velocity of elementary spin excitations is equal to $v$.
It is convenient to characterize the strength of e-e interaction by
the dimensionless coupling constant~\cite{GMP}
$\alpha=(1-K_\rho^2)/(1+3K_\rho^2)$, which in the limit of weak interaction $\alpha \ll 1$
is written as
\begin{equation}
 \alpha\simeq(1-K_\rho)/2\simeq g/2\pi v~.
\label{alpha-small}
\end{equation}

\subsection{Why spin matters}
\label{IIaa}

To qualitatively understand the nature of dephasing of fermionic excitations
in a spinful LL, it is instructive to recall the perturbative
expansion\cite{GMPlett,maslov} of the self-energy of the single-particle
Green's function in the limit of weak interaction $\alpha \ll 1$
and discuss the e-e scattering rate at the Golden-rule level, first in the
absence of disorder. For the spinless case, such an analysis has been
performed in Refs.~\onlinecite{GMPlett,GMP}, and \onlinecite{lehur06}---see also
Refs.~\onlinecite{lehur02,GMPlett,GMP,lehur06} and
Refs.~\onlinecite{GMPlett,lehur06} for a closely related calculation of the
temporal decay of the single-particle Green's function for spinless and
spinful electrons, respectively. Since the physics of dephasing is governed by
inelastic e-e scattering, a natural first step is to calculate at lowest
(second) order in $\alpha$ the e-e scattering rate $\tau_{ee}^{-1}$
given by the imaginary part of the self-energy.

The Golden-rule expression for the e-e collision rate reads
\begin{equation}
\frac{1}{\tau_{ee}^{\rm GR}(\ve)}=\int\! d\omega d\ve' \,
K(\omega)
 \left(f^h_{\ve-\omega}f_{\ve'}
f^h_{\ve'+\omega}+f_{\ve-\omega} f^h_{\ve'}f_{\ve'+\omega}\right),
\label{GR63}
\end{equation}
where
\be
K(\omega)=\eta_s[\,K^H_{++}(\omega)+K^H_{+-}(\omega)\,]+K^F(\omega)
\label{GR65}
\ee
is the kernel of the e-e collision integral and $f_\ve$ is the Fermi
distribution function, $f^h_{\ve}=1-f_{\ve}$. In Eq.~(\ref{GR65}), the
Hartree terms $K^H_{++}\propto g_4^2$ and $K^H_{+-}\propto g_2^2$ are related
to scattering of two electrons from the same $(++)$ or different $(+-)$ chiral
branches, respectively, $K^F=-K^H_{++}$ is the exchange counterpart of
$K^H_{++}$, $\eta_s$ is the spin degeneracy, $\eta_s=1$ for the spinless case
and $\eta_s=2$ for the spinful case.

To order ${\cal O}(\alpha^2)$, the Golden-rule scattering rate and the
self-energy on the mass shell coincide with each other. At the Fermi level
($\ve=0$) we have
\be
{1 \over \tau_{ee}^{\rm GR}(0)} =-2\, \big[\, \eta_s({\rm Im}\Sigma^H_{++}+{\rm
Im}\Sigma^H_{+-})+{\rm Im}\Sigma^F \big]~,
\ee
where the Hartree terms are given by
\bea
{\rm Im}\Sigma^H_{+\pm}&=&-{\pi\over 2}\alpha^2v\!\int \!d\omega
\,\omega\left(\coth{\omega\over 2T}-\tanh{\omega\over 2T}\right)\nonumber \\
&\times&\int \! dq\, \delta(\omega-vq)\delta(\omega\mp vq)
\label{GRfull}
\eea
and the exchange term $\Sigma^F=-\Sigma^H_{++}$.  Peculiar to 1D are highly
singular contributions to $K(\omega)$ related to scattering of electrons
moving in the same direction.  One sees that the contribution of
$\Sigma^H_{++}$ contains a $\delta$-function squared and thus
diverges.\cite{maslov,GMPlett,GMP} The divergency of the perturbative
expression for the probability of scattering of two electrons of the same
chirality simply means that the energy and momentum conservation laws for this
kind of scattering give a single equation $\omega-vq=0$.

For spinless (spin-polarized) electrons, the divergency in ${\rm
Im}\Sigma^H_{++}$ is canceled by the same divergency in the exchange term. The
remaining term ${\rm Im}\Sigma^H_{+-}$ yields\cite{GMPlett,GMP,lehur06}
\be
{1 \over \tau_{ee}^{\rm GR}(0)}= \alpha^2\pi T~.
\label{2.11}
\ee
Note that for scattering of electrons from different chiral branches on each
other, the energy and momentum conservation laws lead to two equalities:
$\omega-vq=0$ and $\omega+vq=0$, which combine to give $\omega,q=0$ for
allowed energy and momentum transfers. This ``quasi-elastic"\cite{GMPlett,GMP}
character of e-e scattering is a peculiarity of 1D: in higher
dimensionalities, the characteristic energy transfer that determines
$\tau_{ee}^{-1}$ in a clean system is of order $T$.

For spinful electrons, the Fock contribution cancels only the part of the
Hartree term ${\rm Im}\Sigma^H_{++}$ that comes from interaction between
electrons with the same spin. The divergent second-order Hartree term that
arises from interaction between electrons with opposite spins remains
uncompensated. This indicates that the main contribution to $\tau_{\rm
ee}^{-1}$ is now related to scattering of electrons from the same chiral
branch. Thus, already the perturbative expansion
demonstrates\cite{maslov,GMPlett,GMP}
a qualitative difference between the cases of spinless and spinful electrons.

In fact, for spinful electrons, the perturbative expansion of $\tau_{\rm
ee}^{-1}$ in powers of $\alpha$ is diverging in the clean limit at each
order. We will analyze the finite-$T$ damping of the single-particle Green's
function for $\alpha\ll 1$ in Sec.~\ref{III}. Here, we stick to the
calculation of $1/\tau^{\rm GR}_{ee}$ within a ``generalized Golden-rule"
scheme. The term ``generalized" means that we go beyond second order in
$\alpha$ by introducing the dynamically screened e-e interaction
$V(\omega,q)$---which is exactly\cite{dzyaloshinskii74} given by the random
phase approximation (RPA) [see Eq.~(\ref{11}) below]. The second
$\delta$-function in the integrand of Eq.~(\ref{GRfull}) comes precisely from
the imaginary part of the retarded propagator ${\rm Im}V(q, \omega)$
if one takes
the propagator at second order in $\alpha$,
\bea
 {\rm Im}V(q, \omega) \simeq -(2\pi \alpha)^2 \omega v [\,\delta(\omega-vq)+\delta(\omega+vq)\,]~.
\nonumber \\
\label{ImVv}
\eea
Using the full RPA propagator
${\rm Im}V(q, \omega) \propto [\,\delta(\omega-uq)+\delta(\omega+uq)\,]$, the
leading at $\alpha\ll 1$ Golden-rule expression for the e-e scattering rate
of spinful electrons is written as
\begin{equation}
\frac{1}{\tau_{ee}^{\rm GR}(0)}\simeq
2\pi\alpha^2 vT\int\! d\omega\!\int \! dq\,
\delta(\omega-vq)\delta(\omega-uq)~,
\label{2.12}
\end{equation}
which only differs from $-2{\rm Im}\Sigma_{++}^H$ in Eq.~(\ref{GRfull}) in
that one of the $\delta$-functions has a shifted velocity $v\to u$. Note that,
similarly to the spinless case, $1/\tau_{ee}^{\rm GR}(0)$ in Eq.~(\ref{2.12}) is
determined by $\omega,q=0$. The relative shift between the arguments of the
$\delta$-functions makes the expression for $1/\tau_{ee}^{\rm GR}(0)$ finite:
\begin{equation}
\frac{1}{\tau_{ee}^{\rm GR}(0)}
\simeq 2\pi\alpha^2 T\,{v\over |u-v|}\simeq   |\alpha|\pi  T~,
\label{tauee-spinfull}
\end{equation}
where we used Eq.~(\ref{2.6}) for $u-v\simeq 2\alpha v$ for small
$\alpha$. Remarkably, the e-e scattering rate for spinful electrons turns out
to be of first order in $\alpha$, in contrast to the spinless case
(\ref{2.11}), where it is of order $\alpha^2$.

As we will see below in a more consistent treatment which does not rely on the
generalized Golden-rule approach, the scattering rate $1/\tau_{ee}^{\rm GR}(0)$
gives a characteristic decay rate for single-particle excitations and also the
characteristic dephasing rate for WL. What the above
consideration teaches us is that in 1D the spin degree of freedom strongly
enhances the e-e scattering rate for weakly interacting electrons. The
parametric difference between the spinless and spinful cases is in stark
contrast to higher dimensionalities, where taking spin into account typically
yields for relaxation rates in a clean system only numerical factors of order
unity.

\subsection{Functional bosonization}
\label{IIb}

The method we use here to study the quantum interference in a disordered
spinful LL is functional bosonization. It was introduced for the clean LL
model in Refs.~\onlinecite{fog76,lee88} and further developed in
Refs.~\onlinecite{yurk02,kopietz,naon,eckern}. In the earlier work\cite{GMPlett,GMP} by
three of us, the functional bosonization framework was extended to deal with
disordered problems and applied to study the transport properties of a
disordered spinless LL. In this subsection we present a brief outline of
the formalism (for more details see Sec.~VII in Ref.~\onlinecite{GMP}), highlighting
the differences between the spinless and spinful cases.

In contrast to ``full bosonization'', conventionally used for a theoretical
description of the LL, the functional bosonization technique preserves both
fermionic (electrons) and bosonic (collective excitations -- plasmons,
spinons) degrees of freedom. This feature of the method is of great advantage
when one has to deal with interacting problems which are most naturally
described in terms of fermionic excitations, e.g., quantum interference
(Refs.~\onlinecite{GMPlett,GMP} and the present work) or
nonequlibrium\cite{gutman08,bagrets08a,bagrets08b} phenomena in a LL. In
particular, the functional bosonization allows for a straightforward treatment
of e-e interaction while residing in the fermionic basis, which is
especially cost-efficient in the disordered case.

The key steps in setting up the formalism for a LL at thermal equilibrium are
as follows:
\begin{itemize}
\item[$\bullet$] a conventional Hubbard-Stratonovich decoupling of the
four-fermion interaction term in the Matsubara action is performed by means of
introducing a bosonic field $\varphi (x, \tau)$;
\item[$\bullet$] interaction of fermions with the field $\varphi$ is
gauged out by means of a local transformation ($\mu=\pm$)
\be
\psi_{\mu\sigma} (x,\tau) \, \rightarrow \, \psi_{\mu\sigma} (x,\tau)
\, {\rm exp} \, [\, i \, \theta_{\mu} (x, \tau) ],
\label{4}
\ee
where the phase $\theta_{\mu} (x, \tau)$ obeys
\be (\partial_\tau-i\,\mu \, v \, \partial_x)\,
\theta_\mu(x,\tau)=\varphi (x,\tau).
\label{5}
\ee
This transformation completely eliminates the coupling between the
fermionic and bosonic fields from the action
(this property is peculiar to 1D);
\item[$\bullet$] upon this transformation, the bosonic part of the action
remains Gaussian. It is this point at which the peculiarity of the LL
model---the exactness of the RPA---comes into play. The correlation
function of the field $\varphi$ is given by
\be
\left<\varphi(x,\tau)\varphi(0,0)\right>=V(x,\tau),
\label{6}
\ee
where $V(x,\tau)$ is the dynamically screened interaction [see
Eq.~(\ref{11}) below];
\item[$\bullet$] an arbitrary time-ordered fermionic average is expressed
through a product of free electron Green's functions and Gaussian averages of
the phase factors $\exp[i\theta_{\mu} (x, \tau)]$ (taken at different
space-time points). The bosonic averages are represented in terms of the
correlation functions
\bea
&&B_{\mu\nu} (x,\tau) = \left< \,[
\theta_\mu (0,0) - \theta_\mu (x,\tau)]\,\theta_\nu (0,0) \,
\right>\nonumber\\
&&
\label{7}
\eea
which are related to the Fourier component $V(q, i\Omega_n)$ of the interaction
propagator (\ref{6}) as
\bea
B_{+\pm}(x,\tau)&=&T\sum_n\int\!  {dq\over
2\pi}\,\left(e^{iqx-i\Omega_n\tau}-1\right) \nonumber \\
&\times& {V(q, i\Omega_n)\over (v q - i\Omega_n)(\pm v q - i\Omega_n)}~,
\label{13} \nonumber  \\
B_{--}(x,\tau) &=& B_{++}(-x,\tau)~, \nonumber \\
B_{-+}(x,\tau) &=& B_{+-}(x,\tau)~.
\eea
Here $\Omega_n=2\pi nT$ is the bosonic Matsubara frequency;
\item[$\bullet$] while calculating observables (closed fermionic loops), e-e
interaction is completely accounted for by attaching the fluctuating gauge
factors to backscattering vertices. If the number of fermionic loops is larger
than one, each of them has to contain at least one pair of backscattering
vertices in order not to be disconnected (Wick's theorem for the
functional bosonization diagrammatic technique).
\end{itemize}

%%%%%%%%%%%%%%%%%%%%%%%%%%%%%%%%%%%%%%%%%%%%%%%

\begin{figure}[ht]
\centerline{
\includegraphics[width=7cm]{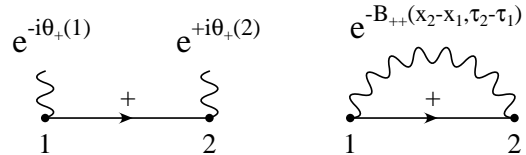}
}
\caption{The Green's function of a right mover propagating between space-time
points $1=(x_1,\tau_1)$ and $2=(x_2,\tau_2)$ before (a) and after (b)
averaging over fluctuations of the gauge factors. Solid line: the bare Green's
function. The wavy lines at the end points represent the factors $\exp [ -i
\theta_{+} (1)]$ and $\exp [ i \theta_{+} (2)]$. The wavy line connecting
points 1 and 2 denotes averaging over fluctuations of $\theta_+(1)$ and
$\theta_+(2)$.}
\label{f13}
\end{figure}

%%%%%%%%%%%%%%%%%%%%%%%%%%%%%%%%%%%%%%%%%%%%%%%%

\begin{figure}[ht] \centerline{
\includegraphics[width=7cm]{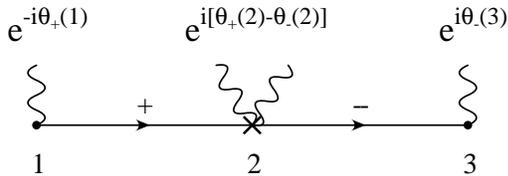} }
\caption{Backscattering of a right mover off an impurity (denoted by a cross)
at point 2. The impurity vertex is dressed by a local gauge factor $\exp
\{i[\theta_+(2)-\theta_-(2)]\}$ which contains two fluctuating fields of
different chirality.}
\label{f14}
\end{figure}

\begin{figure}[ht] \centerline{
\includegraphics[width=6cm]{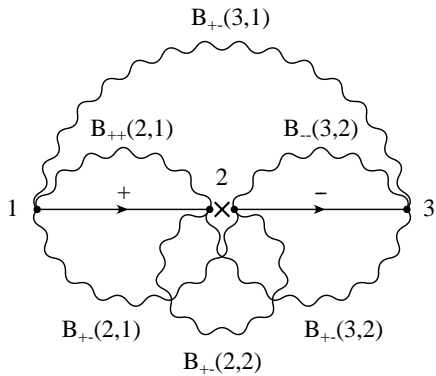} }
\caption{Backscattering off an impurity as shown in Fig.~\ref{f14} after
averaging over fluctuating bosonic fields. Each of the wavy lines represents
a factor of the type $\exp (\pm B_{\mu \nu})$.}
\label{f15}
\end{figure}

As a simple example, consider the single-particle Green's function for, say, a
right mover $G_+(x,\tau)$. Upon gauge transformation (\ref{4}), one gets the
free Green's function $g_+(x,\tau)$ dressed by two phase factors as shown in
Fig.~\ref{f13}(a). Pairing of the two bosonic fields yields
[Fig.~\ref{f13}(b)]
\be
G_{+}(x,\tau)=g_{+}(x,\tau)\,\exp\left[-B_{++}(x,\tau)\right]~.
\label{8}
\ee
More complex quantities are calculated in a similar way.  Each impurity
backscattering vertex at space-time point $N$ generates a phase factor of the
type $\exp \{ \pm i [\theta_{+} (N) - \theta_-(N)]\}$, as illustrated in
Fig.~\ref{f14}. Upon averaging, the phase factors are paired in all possible
ways (Fig.~\ref{f15}). In closed fermionic loops, the correlators
$B_{\mu\nu}(x,\tau)$ [Eq.~(\ref{7})] only appear in the combination
\bea
M(x,\tau)=B_{++}(x,\tau)+B_{--}(x,\tau)-2B_{+-}(x,\tau)~.\nonumber\\
\label{M}
\eea
As a result, each pair of backscattering vertices at points $(x_N,\tau_N)$ and
$(x_{N'},\tau_{N'})$ contributes either the factor
\be
Q(x,\tau)=\exp [M (x, \tau)]~,
\label{Q}
\ee
where $x=x_N-x_{N'}$, $\tau=\tau_N-\tau_{N'}$, or $Q^{-1}(x,\tau)$, depending
on whether chirality of incident electrons at the vertices is the same ($Q$)
or different ($Q^{-1}$).

%%%%%%%%%%%%%%%%%%%%%%%%%%%%%%%%%%%%%%%%%%%%%%%%%

%%%%%%%%%%%%%%%%%%%%%%%%%%%%%%%%%%%%%%%%%%%%%%%%%
%%%%%%%%%%%%%%%%%%%%%%%%%%%%%%%%%%%%%%%%%%%%%%

%%%%%%%%%%%%%%%%%%%%%%%%%%%%%%%%%%%%%%%%%%%%%%%%

The RPA dynamically screened interaction $V(q,i\Omega_n)$ obeys
\be
V^{-1}(q, i\Omega_n)=g^{-1}+\Pi(q, i\Omega_n)~,
\label{9}
\ee
where $\Pi(q,i\Omega_n)$ is the polarization operator. In a spinful LL, the
latter is written as
\be
\Pi(q, i\Omega_n)={2\over \pi v}{v^2q^2\over v^2q^2+\Omega_n^2}~,
\ee
which gives
\be
V(q, i\Omega_n)= g \, \frac{v^2 q^2 + \Omega^2_n}{u^2 q^2 + \Omega^2_n}
\label{11}
\ee
with $u$ from Eq.~(\ref{2.6}).

As will be seen below, it suffices, when calculating the dephasing rate for
weak localization, to deal with the ballistic interaction propagator
(\ref{11}) which does not include backscattering of electrons off
disorder. This should be contrasted with the spinless case, where the
dephasing of localization effects is absent altogether unless the
disorder-induced damping of $V(q, i\Omega_n)$ is taken into
account\cite{GMPlett,GMP} (``dirty RPA").

Substituting Eq.~(\ref{11}) in Eqs.~(\ref{13}) yields
\begin{eqnarray}
B_{++}(x,\tau)&=&-{1\over 2}\ln\eta (x,\tau)-{\alpha_b\over 4}
\ln\varsigma (x,\tau)~,
\nonumber \\
B_{+-}(x,\tau)&=&-{\alpha_r\over 4} \ln\varsigma (x,\tau)~,
\label{14}
\end{eqnarray}
where
\begin{eqnarray}
\varsigma (x,\tau)&=&{(\pi T/\Lambda)^2\over \sinh [\,\pi T(x/u+i\tau)\,]
\sinh [\,\pi T(x/u-i\tau)\,]}, \nonumber \\
\eta (x,\tau)&=&{v \over u}{\sinh[\pi T(x/v  + i\tau)]\over
\sinh [\pi T(x/u + i\tau)]}
\label{15}
\end{eqnarray}
and the constants $\alpha_b$ and $\alpha_r$ are given by
\be
\alpha_b=\frac{(u-v)^2}{2 u v} ~,\quad \alpha_r=\frac{u^2-v^2}{2 u v}~.
\label{16}
\ee
Inspecting Eqs.~(\ref{14}) and (\ref{15}), we see that there is an extra factor of
1/2 in front of both $B_{++}$ and $B_{+-}$ as compared to the
spinless\cite{GMP} case. It is this factor that is responsible for the SCS.

The exponents $\alpha_b$ and $\alpha_e=\alpha_r+\alpha_b$ determine a
power-law suppression of the tunneling density of states (zero-bias anomaly)
for tunneling in the bulk and in the end of a LL, respectively (see, e.g.,
Refs.~\onlinecite{sol79,voit94,schulz95,gog98,schulz00,maslov,giam04}).
In the rest of the paper, we will treat the interaction strength
$\alpha$ as a small parameter. The hierarchy of the constants (\ref{16}) is
then as follows
\be
\alpha_b \ll \alpha_r\ll 1~,
\ee
since $\alpha_b$ is quadratic in $\alpha$, whereas $\alpha_r$ is linear.
Using this hierarchy will greatly simplify the calculation below.

\section{Single-Particle Spectral Properties and Spin-Charge Separation}
\label{III}
\setcounter{equation}{0}

In this section, we use the functional bosonization formalism to study the
single-particle spectral properties of electrons at finite temperature. We begin with
the space-time representation in Sec.~\ref{IIIa}. Then we transform to a
``mixed" space-energy (Sec.~\ref{IIIb}) and the momentum-energy
(Sec.~\ref{IIIc}) representations, using approximations appropriate in the
weak-interaction limit $\alpha\ll 1$. The analysis of the various
representations of the single-particle Green's function will serve as a
starting point for the calculation of the WL and ME terms in the conductivity
in Secs.~\ref{IV} and \ref{V}.

\subsection{Green's function in the $(x,\tau)$ representation:\\
Weak-interaction approximation}
\label{IIIa}

\begin{figure}[ht]
\centerline{ \includegraphics[width=8cm]{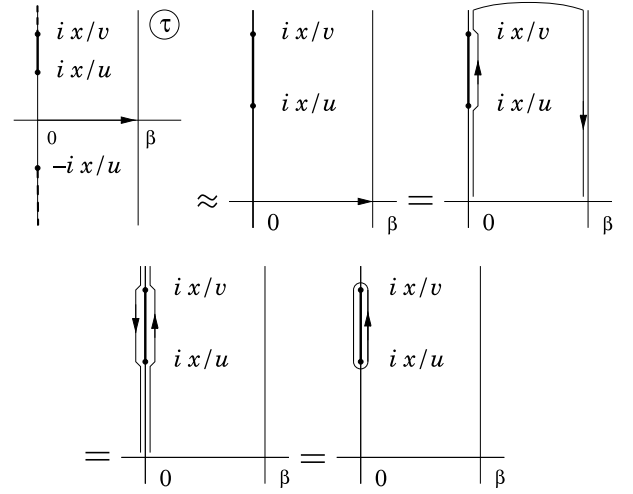} }
\caption{Left top panel: the analytical structure of the
right-mover Green's function $G_{+} (x, \tau)$ in the complex plane of the
Matsubara time $\tau$. The bold solid and the dashed lines represent
``strong'' and ``weak" branch cuts, respectively [see the text below
Eq.~(\ref{18})]. Central top panel: within the approximation (\ref{19}), only
the square-root branch cut between the points $\tau=ix/u$ and $ix/v$
survives. Last three panels: the contour transformation used to calculate the
Green's function in the space-energy representation in Sec.~\ref{IIIb}. }
\label{complex}
\end{figure}

In the absence of interaction, the single-particle Green's function of right
(+) and left $(-)$
movers $g_\pm (x,\tau)$ is given by
\be
g_\pm (x,\tau)=\mp {i T \over 2 v }\,{1\over \sinh [\pi T(x/v \pm i\tau)]}~.
\label{17}
\ee
Plugging Eqs.~(\ref{14}), (\ref{15}), and (\ref{17}) into Eq.~(\ref{8}), we
have for the Green's function of right movers in a spinful LL:
\bea
&&G_{+}(x,\tau)= - \frac{i}{2 \pi \sqrt{u v}} \nonumber\\ &&\times\left\{
\frac{\pi T}{\sinh [\pi T(x/v + i\tau)]} \frac{\pi T}{\sinh [\pi T(x/u +
i\tau)]}\right\}^{1/2} \nonumber\\ &&\times\left\{ \frac{\pi T /
\Lambda}{\sinh [\pi T (x/u + i\tau)]} \frac{\pi T / \Lambda}{\sinh [\pi T(x/u
- i\tau)]}\right\}^{\alpha_b/4}~, \nonumber\\
\label{18}
\eea
in agreement with the result obtained by conventional bosonization (see, e.g.,
Refs.~\onlinecite{voit94,schulz95,gog98,schulz00,maslov,giam04}) and by purely
fermionic methods (see, e.g., Ref.~\onlinecite{sol79}). The Green's function
of left movers $G_-(x,\tau)=G_+(-x,\tau)$. The analytical structure of
$G_+(x,\tau)$ in the complex plane of $\tau$ ($0<{\rm Re}\,\tau<1/T$) is
shown in Fig.~\ref{complex} for positive $x$. There are three branch points
at $\tau= i x/u$, $i x/ v$, and $- i x/u$. One way to choose branch cuts is
shown in the top left panel of Fig.~\ref{complex}: those starting at
$\tau=ix/u$ and $ix/v$ are sent upwards, whereas that starting at $\tau=-ix/u$
is sent downwards. If $\alpha\ll 1$, the first two cuts are much different
from the third one. In the limit of small $\alpha$, the cut that connects the
points $\tau= i x/u$ and $i x/ v$ corresponds to an almost square-root
singularity, so that the main change the Green's function experiences when
crossing this cut is a change of sign (``strong cut"). On the other hand, the
cut that goes from $- i x/u$ to $-i\infty$ is ``weak'' in the sense that the
discontinuity of $G_{+}(x, \tau)$ across this cut is proportional to $\alpha_b
\sim \alpha^2 \ll 1$. Similarly, crossing the axis of imaginary $\tau$ between
$\tau=ix/v$ and $i\infty$ is associated with a weak discontinuity.

The main approximation we make in this paper consists in sending $\alpha_b$ to
zero everywhere in the calculation while keeping the effects of leading
(linear) order in the interaction strength. That is, below we retain the
difference between $u$ and $v$, Eq.~(\ref{2.6}),
\be
u\simeq v(1+2\alpha)~,\quad\alpha\ll 1~,
\ee
as the only effect of e-e interaction.~\cite{footg2}
The Green's function $G_{+}(x,\tau)$ then reads
\bea
&&G_{+}(x,\tau)\simeq  - \frac{i}{2 \pi \sqrt{u v}}  \nonumber\\
&&\times  \left\{\frac{\pi T}{\sinh [\pi T(x/v + i\tau)]}
\frac{\pi T}{\sinh [\pi T(x/u + i\tau)]} \right\}^{1/2}~.\nonumber\\
\label{19}
\eea
The velocities $u$ and $v$ in Eq.~(\ref{19}) coincide with the velocities of
the elementary collective excitations (plasmons and spinons). The appearance
of the two velocities in the single-particle correlator signifies SCS. Within
the approximation (\ref{19}), the two velocities enter the fermionic Green's
function in a symmetric way. The analytical structure of $G_+(x,\tau)$ in
Eq.~(\ref{19}) is simplified to a single square-root cut between the points
$\tau=ix/u$ and $ix/v$, as illustrated in the center top panel of
Fig.~\ref{complex}. We will use the approximation (\ref{19}), which captures
the essential physics of SCS, throughout the paper below.

%%%%%%%%%%%%%%%%%%%%%%%%%%%%%%%%%%%%%%%%%%%%%%%%%%%%%%

%%%%%%%%%%%%%%%%%%%%%%%%%%%%%%%%%%%%%%%%%%%%%%%%%%%%%%%

It is instructive to compare Eq.~(\ref{19}) with the Green's function of
spinless electrons $G_{+}^{sl} (x,\tau)$ (see, e.g.,
Refs.~\onlinecite{sol79,voit94,schulz95,gog98,schulz00,giam04,maslov,GMP}).
Since the correlator $B_{++}(x,\tau)$ in the spinless case is twice as large,
$G_+^{sl}(x,\tau)$ is given by
\bea
&&G_{+}^{sl} (x,\tau)= - \frac{i}{2 \pi u}  \,
\frac{\pi T}{\sinh [\pi T(x/u + i\tau)]}  \nonumber\\
&&\times  \left\{ \frac{\pi T / \Lambda}{\sinh [\pi T (x/u + i\tau)]}
\frac{\pi T / \Lambda}{\sinh [\pi T(x/u -
  i\tau)]}\right\}^{\alpha_b/2}~. \nonumber\\
\label{18a}
\eea
One sees that in the absence of spin, interaction leads to a replacement of
the velocity $v \to u$ in the bare Green's function and generates two brunch
cuts: one with an exponent close to 1 and the other with the small exponent
$\alpha_b/2$. It follows that for spinless electrons the approximation
$\alpha_b\to 0$, analogous to Eq.~(\ref{19}), would eliminate all dephasing
effects---since the latter only originate from the factors in the second line
of Eq.~(\ref{18a}). By contrast, dephasing in the spinful case arises already
at order ${\cal O}(\alpha)$, as will be shown below. Consequently, the
approximation (\ref{19}) allows us to obtain---in a controllable
way---analytical results valid in the limit $\alpha\ll 1$.

Let us now identify two important spatial scales.
For this purpose, we perform the
Wick rotation $\tau \to i (t+i0)$ in Eq.~(\ref{19}).
For large $|x/ u- t|,|x/v - t|\gg 1/T$, Eq.~(\ref{19}) yields
\be
G_{+}(x,it)\propto \exp\left[ -\pi T \left( \left|x/u -t \right| +
\left| x/v - t \right|    \right) /2 \right]~.
\label{20}
\ee
Within the interval $x/u < t < x/v$, the Green's function given by
Eq.~(\ref{20}) decays as
\be
G_{+} (x,it) \propto \exp \left(- x/2l_{ee} \right)~,
\label{21}
\ee
independently of $t$, whereas outside this interval the Green's function
is suppressed much more strongly; in particular, at $t=0$:
\be
G_{+} (x,0) \propto \exp \left( - x/2l_{T} \right)~,
\label{22}
\ee
merely due to the thermal smearing.
Here we have introduced
\be
l_{ee}= \frac{u_-}{2\pi T}
\label{23}
\ee
and
\be
l_T = \frac{u_+}{2\pi T}~,
\label{23a}
\ee
which are the length scale of spatial decay of fermionic excitations due to
e-e interaction and the ``thermal smearing length", respectively.  The
velocities $u_\pm$ are given by
\be
{1\over u_\pm}=\frac{1}{2}\left( {1\over v}\pm {1\over u}\right)~.
\label{24}
\ee
The length $l_{ee}$ has also been termed the Aharonov-Bohm dephasing
length.\cite{GMPlett,GMP,lehur02,lehur06} Note that this length agrees
with the Golden-rule estimate (\ref{tauee-spinfull}), up to a numerical factor.

For $\alpha\ll 1$ we have 
\be
l_T\simeq v/2\pi T, \quad  l_{ee} \simeq l_T/\alpha,
\label{lTee-small-alpha}
\ee 
i.e.,
$l_{ee}$ for weak interaction is much longer than $l_T$. In Secs.~\ref{IV} and
{V}, when considering the system in the presence of disorder, there will
appear one more characteristic length scale: the electron mean free path due
to backscattering off impurities $l$. We will assume that $T$ is sufficiently
large, so that $l_{ee}\ll l$. As will be seen below, this condition means that
the disordered system is in the WL regime. For lower temperatures, strong
localization sets in.\cite{SL} Altogether, the hierarchy of length scales
in our problem is
\be
l_T\ll l_{ee}\ll l~.
\label{25}
\ee

\subsection{Green's function in the $(x,\ve)$ representation}
\label{IIIb}

We now turn to the single-particle Green's function in the space-energy
representation, which is obtained by Fourier-transforming $G_{+}(x,\tau)$ with
respect to $\tau$. Within the
small-$\alpha$ approximation (\ref{19}), the only singularity of
$G_{+}(x,\tau)$ is a branch cut between $\tau=|x|/u$ and $|x|/v$ in the upper
or lower half-plane of $\tau$ depending on the sign of $x$. Since
$G_+(x,\tau)$ in this approximation is analytical in one of the half-planes of
$\tau$, its Fourier transform vanishes for $\ve_n<0$ if
$x>0$ or for $\ve_n>0$ if $x<0$. For both $\ve_n$ and $x$ positive, we
transform the contour of integration as shown in Fig.~\ref{complex}.  Closing
the contour upwards, the integral along the real axis of $\tau$ is represented
as a sum of two integrals along the imaginary axis at $\tau=+0$ and
$\tau=1/T-0$. In view of the periodicity of $G_+(x,\tau)$ in $\tau$, the sum
gives the integral along the contour around the cut. Closing similarly the
contour of integration downward if both $\ve_n$ and $x$ are negative,
we get
\be
\int_{0}^{1/T}\!d\tau\,\exp (i\ve_{n}\tau )\,
G_+(x,\tau)=G_+^r(x,i\ve_n)- G_+^a(x,i\ve_n)~.
\label{3.14}
\ee
Here $\ve_n=2\pi (n+\frac{1}{2})T$ is the fermionic Matsubara frequency,
\bea
G_+^r(x,i\ve_n)&=& \theta (\ve_n)\theta(x){\mathscr G} (x,\ve_n)~,
\nonumber \\
G_+^a(x,i\ve_n)&=&\theta (-\ve_n)\theta(-x){\mathscr G}(x,\ve_n)~,
\label{27}
\eea
and the function ${\mathscr G}(x,\ve_{n})$ depends on the absolute values of the
coordinate and energy:
\bea
{\mathscr G}(x,\ve_{n})&=&\frac{T}{i\sqrt{uv}} \,\, \exp (-
|\ve_{n}x| / u ) \int_{0}^{2|x|/u_-} \!\!\!  dt  \nonumber  \\
&\times&\frac{ \exp (-|\ve_{n}| t)}{[\,\sinh (\pi Tt)\,
\sinh (2 \pi T |x| /u_--\pi Tt)\,]^{1/2}}~.  \nonumber\\
\label{28}
\eea
For
left movers, $G^{r,a}_-(x,i\ve_n)=G^{r,a}_+(-x,i\ve_n)$.
Integration in Eq.~(\ref{28}) yields (in the rest of the subsection, let both
$\ve_n$ and $x$ be positive):
\bea
{\mathscr G} (x,\ve_{n})&=&\frac{\exp\left( 
-\ve_n x /u + x/ 2 l_{ee}\right)}{i\sqrt{u v}} \nonumber \\
&\times& _{2} {\rm F}_{ 1} \left[\, 1/2 + \xi_n , \, 1/2, \, 1; \,
  \chi (x)  \,\right]~,
\label{29}
\eea
where $_{2}{\rm F}_{ 1} (a, b, c ; z)$
is the hypergeometric function,
\be
\chi (x)= 1 - \exp (2 x / l_{ee}), \qquad \xi_n= \ve_n/2 \pi T~.
\label{30}
\ee

We now analyze the asymptotic behavior of ${\mathscr G} (x,\ve_{n})$ as a function
of two dimensionless parameters $x/l_{ee}$ and $\ve_n/2\pi T$. For
$x/l_{ee}\gg 1$ and $\ve_n/2\pi T\geq 1$, Eq.~(\ref{29}) gives
\bea
{\mathscr G} (x \gg l_{ee}, \ve_n) &\simeq&
  \frac{\exp\left(-\ve_n x /u -  x / 2 l_{ee}\right)}{i \sqrt{\pi} \, \sqrt{ u
      v}} \,
\frac{\Gamma (\xi_n)}{\Gamma
  (\frac{1}{2} + \xi_n )}~, \nonumber \\
\label{31}
\eea
where $\Gamma(z)$ is the gamma-function.  After the analytical continuation to
real energies $i\ve_n\to \ve+i0$, Eq.~(\ref{31}) reveals oscillations of $
G_+^r(x,\ve)$ as a function of $\ve x/u$ and an exponential decay as a function
of $x/l_{ee}$. Using Eq.~(\ref{31}) for the analytical continuation is only
accurate for $(|\ve|/T)(x/l_{ee})\gg 1$. In the opposite limit, one has to
analytically continue already in Eq.~(\ref{29}), which yields the ``static
limit" for the Green's function with
\be
{\mathscr G}(x,0)\simeq \frac{2 \exp (x / 2 l_{ee})}
{i\pi \sqrt{u v}}\,{\rm K} \big[ \chi (x) \big]~,
\label{33}
\ee
where ${\rm K}(z)$ is the complete elliptic integral. For $x/l_{ee}\gg 1$,
Eq.~(\ref{33}) reduces to
\be
{\mathscr G}(x\gg l_{ee},0) \simeq\frac{2}{i\pi \sqrt{u v}}\,
\frac{x}{ l_{ee}}\exp (-x / 2 l_{ee})~.
\label{32}
\ee
Continued to real energies, Eqs.~(\ref{31}) and (\ref{32}) match onto each
other at $(|\ve|/T)(x/l_{ee}) \sim 1$. Finally, the high-energy
short-distance asymptotic behavior of the Green's function is given by
\bea
{\mathscr G} (x \ll l_{ee}, \ve_n \gg T) \simeq \frac{\exp ( - \ve_n
    x/u_{+})}{i\sqrt{u v}} \,\,  I_{0} \left( \frac{\ve_n
    x}{ u_{-}}  \right)~,\nonumber\\
\label{35}
\eea
where $I_0(z)$ is the Bessel function of the imaginary argument.

\begin{figure}[ht]
\centerline{\includegraphics[width=8cm]{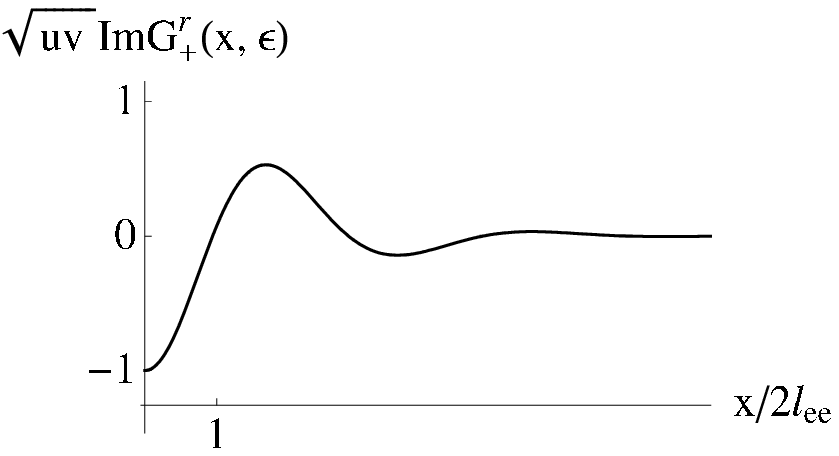}}
\caption{Imaginary part of the Green's function
$G_{+}^r (x,\ve)$ [obtained as an analytical continuation of Eq.~(\ref{28})
onto the real axis of $\ve$ from above] as a function of $x$ for small energies $\ve\ll T$ shows
oscillations $\cos (x\ve/u) \exp(-x/2l_{ee})$ with a period $2\pi u/\ve$,
exponentially suppressed on the scale of $l_{ee}$. The parameters of the plot are:
$\ve/T=0.25$ and $u/v=1.1$.}
\label{greenl}
\end{figure}

\begin{figure}[ht]
\centerline{
\includegraphics[width=8cm]{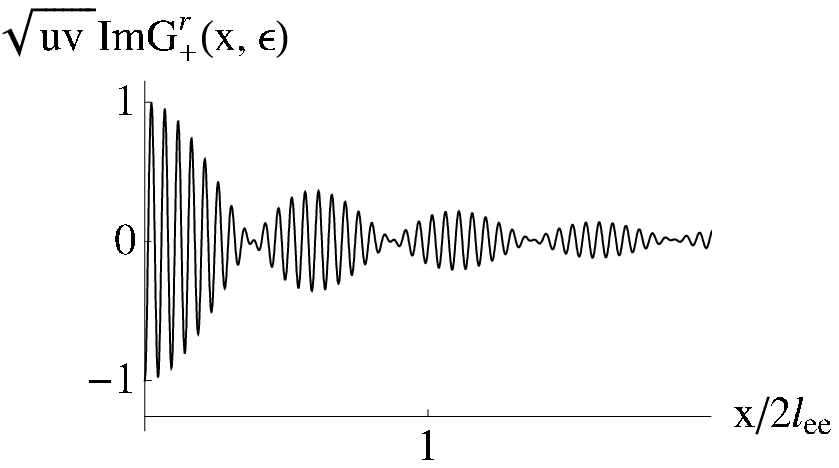}
}
\caption{Imaginary part of the Green's function $G_{+}^r (x,\ve)$ as a function
of $x$ for large energies $\ve \gg  T$.  One sees oscillations with a
period $2\pi u_+/\ve$, beatings with a period $2\pi u_-/\ve$, and a power-law
decay $x^{-1/2}$ (taking place up to $x\sim l_{ee}$). The asymptotic behavior
for large $x\gg l_{ee}$ at which $G_+^r(x,\ve)$ is suppressed exponentially is
not shown in the figure. The parameters of the plot are:
$\ve/T=20$ and $u/v=1.1$. }
\label{greenh}
\end{figure}

The imaginary part of $G^r_{+} (x,\ve)$ [obtained as an analytical continuation of
$G^r_+(x,i\ve_n)$ onto the real axis of energy from the upper half-plane]
as a function of $x$ for small and
large $\ve / T$ is shown in Figs.~\ref{greenl} and \ref{greenh},
respectively.  While in the former case there are only simple oscillations
which are suppressed exponentially on the scale of $l_{ee}$ [cf.\
Eq.~(\ref{31})], the behavior of $G_{+}^r (x,\ve)$ in the latter case is
richer. Specifically, Fig.~\ref{greenh} exhibits beatings and an intermediate
power-law decay, in agreement with Eq.~(\ref{35}). The real part of $
G_+^r(x,\ve)$ behaves similarly.

%%%%%%%%%%%%%%%%%%%%%%%%%%%%%%%%%%%%%%%%%%%%%%%%%%%%

%%%%%%%%%%%%%%%%%%%%%%%%%%%%%%%%%%%%%%%%%%%%%%%%%%%%

\subsection{Green's function in the $(q, \ve)$ representation:\\ Spectral
weight}
\label{IIIc}

Here, we complete the analysis of the single-particle Green's function in a
spinful LL by inspecting its spectral properties in the momentum-energy
representation. Fourier-transforming Eq.~(\ref{3.14}) with respect to $x$, 
and analytically continuing the result onto the real axis of $\ve$ 
from the upper half-plane, $i \ve_n \to \ve
+ i 0$, we get the retarded Green's function $G_{+}^{R} (q,\ve)$ in 
the $(q, \ve)$ representation.
Similarly, the analytical continuation onto the real axis from below, 
$i\ve_n\to\ve-i0$, yields the advanced Green's function $G_{+}^{A} (q,\ve)$.

The retarded and advanced Green's functions of right movers can be written
in the form
\be
G_{+}^{R,A} (q,\ve) = \frac{2 \, l_{ee}}{\sqrt{u v}} {\cal P}(\pm
\kappa_{u}) {\cal P}(\pm \kappa_{v})\ ,
\label{40}
\ee
where the signs $+$ and $-$ correspond to the retarded ($R$) and advanced ($A$) 
Green's functions, 
respectively,
\be
\kappa_u =(\ve/u-q)l_{ee}~, \quad \kappa_v =(\ve/v-q)l_{ee}~,
\ee
and
\be
{\cal P} (z) = \frac{\Gamma\left[(1- 2 \, i z)/4
  \right]}{\Gamma\left[(3- 2 \, i z)/4 \right]}~.
\label{37}
\ee
As a function of complex variable $z$, ${\cal P} (z)$ has a series of simple
poles (originating from the gamma-function in the numerator) at $z=-i(4m+1)/2$, 
where $m$ is a positive integer. The pole that is closest to the real
axis [corresponding to complex $\ve=uq-iu/2l_{ee}$ and $\ve=vq-iv/2l_{ee}$ in Eq.~(\ref{40})] 
determines the spatial/temporal decay of the Green's functions considered 
in Secs.~\ref{IIIa} and \ref{IIIb}.

An alternative representation, which straightforwardly splits $G_{+}^{R,A}
(q,\ve)$ into the real and imaginary parts, is
\bea
G_{+}^{R,A} (q,\ve)&=&\frac{ l_{ee}}{\sqrt{u v }}
{\cal L}^{R,A}(\kappa_u, \kappa_v) {\cal K} (\kappa_u) {\cal K} (\kappa_v),
\nonumber\\
\label{41}
\eea
where
\bea
{\cal L}^{R,A}(x, y) &=&  \sinh\left[\,\pi (x+y)/2 \,\right]
\nonumber \\
 &\mp&  i \cosh\left[\,\pi (x-y)/2\,\right]
\label{42}
\eea
and the real function ${\cal K}(z)$ is given by
\be
{\cal K}(z) = \frac{1}{2\pi}\Gamma\left[(1-  2 \, i z)/4 \right]\Gamma\left[(1+ 2 \, i z)/4 \right]~.
\label{43}
\ee
The upper and lower signs in Eqs.~(\ref{41})--(\ref{43}) correspond to the
retarded and advanced functions, respectively.

Note that the influence of interaction on the behavior of $G_{+}^{R,A}
(q,\ve)$ is twofold. Firstly, it factorizes the single-particle fermionic
Green's function into two parts characterized by different velocities---$v$
for the spin factor and $u$ for the charge factor. This is the essence of
the SCS. Secondly, at finite $T$, interaction leads to a broadening of the
singularities in the spectral weight, i.e., to a shift of the singularities of
$G_+^{R,A}(q,\ve)$ into the complex plane. In the ($x,\ve$) representation,
this shift manifests itself in the exponential damping of the Green's function
on the spatial scale of $l_{ee}$, as discussed in Sec.~\ref{IIIb}.

\begin{figure}[ht]
\centerline{
\includegraphics[width=8cm]{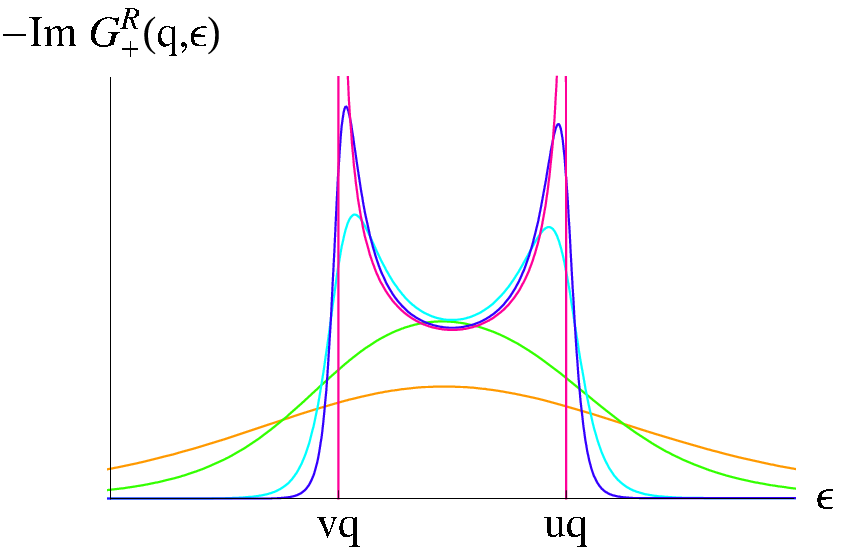}
}
\caption{
Spectral weight ${\rm Im}G_+^R(q,\ve )$ (arbitrary units) for right movers as a function of
energy for different temperatures and $u/v=1.1$. At low $T$, the SCS
manifests itself in the double-peak structure, with two peaks located at $\ve=
v q$ and $\ve = uq$.  When $T$ increases ($T/vq=0,\ 0.05,\ 0.1,\ 0.5$ and $1$)
the singularities are rounded off and
eventually a single peak emerges.
}
\label{imaginary}
\end{figure}

\begin{figure}[ht]
\centerline{
\includegraphics[width=8cm]{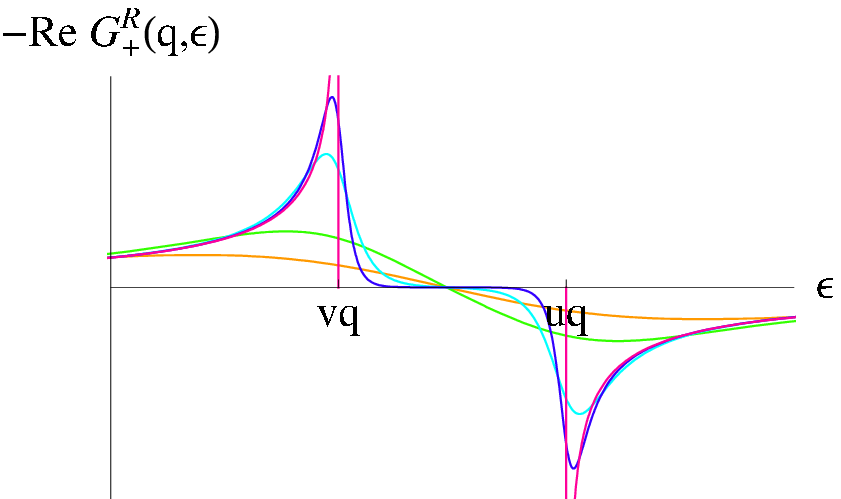}
}
\caption{
Real part of the right-mover Green's function $G_+^R(q,\ve)$ as a function of
energy. Parameters are the same as in Fig.~\ref{imaginary}. The singularities
are smoothened and the peak-dip structure broadened with increasing
temperature.
}
\label{real}
\end{figure}

Figures \ref{imaginary} and \ref{real} illustrate how the imaginary and real
parts of $G_+^R(q,\ve)$ as a function of $\ve$ evolve with varying temperature.
At $T = 0$ one gets:\cite{voit93}
\be
G^R_{+} (q, \ve)  = 
{1 \over \sqrt{\ve - uq } \, \sqrt{  \ve - vq }}
\label{GT0}
\ee
(for $u\to v$, $\ve$ is understood as $\ve+i0$).
At low
$T$, there is a double-peak structure which represents the SCS, with
square-root singularities at $\ve = v q$ and $\ve = u q$, weakly smeared by
temperature. With increasing $T$, the broadening becomes more pronounced and
eventually two peaks in the spectral weight merge into a single peak of width
$\sim \alpha T$.

At this point, it is worth recalling that we have neglected effects of
interaction which are related to the exponent $\alpha_b\sim {\cal
O}(\alpha^2)$ in Eq.~(\ref{18}). Retaining $\alpha_b$, i.e., including the
last factor in Eq.~(\ref{18}) would only lead to the following two effects,
both of which are of minor importance in our consideration at weak
interaction.  Firstly, there will be an additional small asymmetry between two
peaks in Figs.~\ref{imaginary} and \ref{real}. Secondly, an additional, weak
singularity (characterized by the exponent $\alpha_b$) will arise at $\ve = -
u q$ (cf.\ Refs.~\onlinecite{meden92} and \onlinecite{voit93}, where the single-particle
Green's function in a spinful LL was investigated at $T=0$ beyond the weak-interaction
limit).

%%%%%%%%%%%%%%%%%%%%%%%%%%%%%%%%%%%%%%%%%%%%%%%%%%%%%%%%%%

%%%%%%%%%%%%%%%%%%%%%%%%%%%%%%%%%%%%%%%%%%%%%%%%%%%%%%%%%%

%%%%%%%%%%%%%%%%%%%%%%%%%%%%%%%%%%%%%%%%%%%%%%%%%%%%%%%%%%%

%%%%%%%%%%%%%%%%%%%%%%%%%%%%%%%%%%%%%%%%%%%%%%%%%%%%%%%%%%%%%%

To conclude, in Sec.~\ref{III} we have analyzed the behavior of
single-particle excitations in a spinful LL at finite $T$. We have
demonstrated that, because of the SCS, it is dramatically modified as compared
to the spinless case. In particular, the decay length $l_{ee}$ turned out to
be parametrically shorter than for spinless electrons. However, {\it a priori}
it is not immediately clear to what extent the modification of the single-particle
properties will affect the transport (i.e., two-particle for fermions)
properties of a spinful LL. Indeed, as shown in Refs.~\onlinecite{GMPlett,GMP}
for the spinless case, the WL dephasing length $l_{\phi}$ is parametrically longer
than $l_{ee}$. Calculation of the conductivity of a disordered quantum wire in the
presence of spin is a subject of the next section.

\section{Weak Localization}
\label{IV}
\setcounter{equation}{0}

So far we have analyzed the single-particle properties of a spinful LL
at finite $T$ in the absence of disorder. Now we introduce disorder and turn to the
calculation of a two-particle quantity, namely, the conductivity.
At high $T$, the leading term in the conductivity is given by the Drude formula,
\begin{equation}
 \sigma_{\rm D}=\frac{2e^2}{\pi}l,
\label{sigmaD}
\end{equation}
with a renormalized~\cite{mattis74,luther74,giam88,kane92}
by interaction---therefore temperature dependent---mean free
path $l\propto T^{\alpha_r}$.

We consider now a correction $\Delta\sigma$ to the Drude conductivity,
associated with the quantum interference of electron
waves multiple-scattered off disorder. In 1D, the leading contribution to
$\Delta\sigma$ comes from a Cooperon-type scattering process which, in
contrast to higher dimensionalities, involves a minimal possible number of
scatterings on impurities, namely three (``three-impurity
Cooperon'').\cite{GMPlett,GMP} The peculiarity of 1D in this respect is that a
single-channel quantum wire is in the WL regime---not strongly
localized---only if the dephasing length $l_\phi$ that cuts off the WL
correction is shorter than the mean free path $l$. That is, the WL correction
is accumulated on ballistic scales---hence the shortest possible Cooperon
ladder with three impurity legs.

\subsection{General expression for Cooperon}
\label{IVa}

\begin{figure}
\centerline{
\includegraphics[width=8cm]{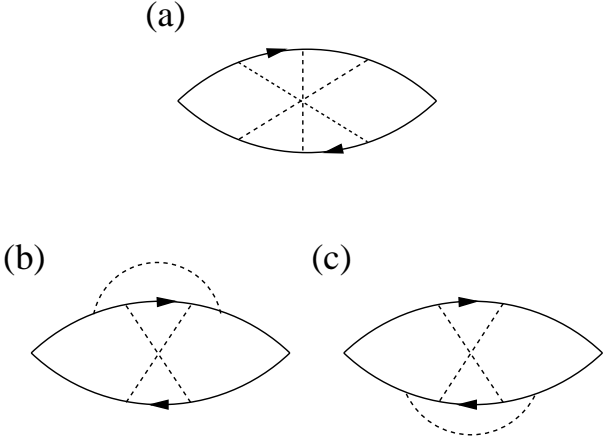}
}
\caption{
Diagrams giving the leading contribution to the interference correction to the
conductivity. The dashed lines represent the impurity scatterings and the
solid lines denote the electron Green's functions (with the disorder effects
incorporated at the self-energy level) and the dashed lines represent the
impurity-induced backscattering. The diagrams are understood as ``dressed''
by interaction as shown in Fig.~\ref{diagram}.
}
\label{3diagrams}
\end{figure}

\begin{figure}[ht]
\centerline{
\includegraphics[width=4.5cm]{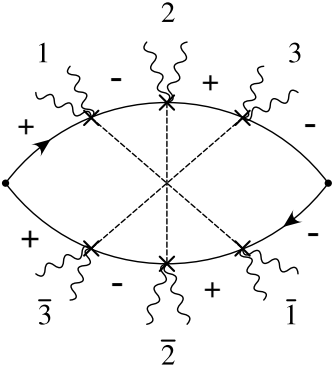}}
\caption{
The same diagram as in Fig.~\ref{3diagrams}a with the interaction-induced
factors $\exp (\pm i\theta_\mu)$ shown explicitly. The solid lines with arrows
stand for bare Green's functions, the crosses for the impurity vertices,
and the wavy lines for the factors $\exp (\pm i\theta_\mu)$. The space-time
coordinates of the backscattering vertices are denoted by $N = (x_N, \tau_N)$
and ${\bar N} = (x_N, {\bar \tau}_N)$. Averaging over the fields $\theta_\mu$
couples all wavy lines with each other, cf.\ Fig.~\ref{f15}.
}
\label{diagram}
\end{figure}

The leading term in $\Delta\sigma$ is given by the
diagrams\cite{GMPlett,GMP} in Fig.~\ref{3diagrams}. These are understood as
dressed by interaction-induced fluctuating gauge factors $\exp [\pm i
\theta_\mu(x,\tau)]$ attached pairwise to the backscattering vertices, as
described in Sec.~\ref{IIb} and illustrated in Fig.~\ref{diagram} for the case
of diagram (a). The sum of contributions of diagrams (b) and (c) is equal to
that of diagram (a). At this level, there is no difference in the structure of the diagrams
between the spinful
and spinless cases---the only difference stems from the particular form of the
correlators of the phases $\theta_\mu(x,\tau)$.

%%%%%%%%%%%%%%%%%%%%%%%%%%%%%%%%%%%%%%%%%%%%%%%%%%%%%%%%%%%%%%%

%%%%%%%%%%%%%%%%%%%%%%%%%%%%%%%%%%%%%%%%%%%%%%%%%%%%%%%%%

%%%%%%%%%%%%%%%%%%%%%%%%%%%%%%%%%%%%%%%%%%%%%%%%%%%%%%%%%

Averaging over the fields $\theta_\mu(x,\tau)$, we get for the interefernce correction
at Matsubara frequency $\Omega_m$:
\begin{widetext}
\bea
\Delta \sigma(i\Omega_m)&=&2 \times  2 \times  2 \times  ( e  v  )^2 \times
\left({v^2 \over 2  l_0} \right)^3  \times   {1 \over \Omega_m}
\nonumber \\
& \times &
\frac{T}{L} \int_0^{1/T} \!d \tau_1
\int_0^{1/T} \!d \bar{\tau}_1
\int_0^{1/T} \!d \tau_2
\int_0^{1/T} \!d \bar{\tau}_2
\int_0^{1/T} \!d \tau_3
\int_0^{1/T} \!d \bar{\tau}_3
\,
\int \!d x_1
\int \!d x_2
\int \!d x_3
\nonumber \\
&\times&
\left[g_+(x_1-x_3,\tau_1-\bar{\tau}_3)
Q^{-1}(x_1-x_3,\tau_1-\bar{\tau}_3)\right]\,
\left[g_-(x_2-x_1,\tau_2-\tau_1)
Q^{-1}(x_2-x_1,\tau_2-\tau_1)\right]
\nonumber\\
&\times&
\left[g_+(x_3-x_2,\tau_3-\tau_2)
Q^{-1}(x_3-x_2,\tau_3-\tau_2 )\right]\,
\left[g_-(x_1-x_3,\bar{\tau}_1-\tau_3)
Q^{-1}(x_1-x_3,\bar{\tau}_1-\tau_3)\right]
\nonumber\\
&\times&
\left[g_+(x_2-x_1,\bar{\tau}_2-\bar{\tau}_1)
Q^{-1}(x_2-x_1,\bar{\tau}_2-\bar{\tau}_1)\right]\,
\left[g_-(x_3-x_2,\bar{\tau}_3-\bar{\tau}_2)
Q^{-1}(x_3-x_2,\bar{\tau}_3-\bar{\tau}_2)\right]
\nonumber\\
&\times&
Q(x_1-x_3,\tau_1-\tau_3)\,
Q(x_1-x_3,\bar{\tau}_1-\bar{\tau}_3)\,
Q(x_1-x_2,\tau_1-\bar{\tau}_2)
\nonumber\\
&\times&
Q(x_2-x_1,\tau_2-\bar{\tau}_1)\,
Q(x_3-x_2,\tau_3-\bar{\tau}_2)\,
Q(x_2-x_3,\tau_2-\bar{\tau}_3)
\nonumber\\
&\times&
Q^{-1}(0,\tau_1-\bar{\tau}_1)\,
Q^{-1}(0,\tau_2-\bar{\tau}_2)\,
Q^{-1}(0,\tau_3-\bar{\tau}_3)
\nonumber\\
&\times&
{\cal W}_+^i (x_1-x_3,\tau_1,\bar{\tau}_3,\Omega_m)
{\cal W}_-^f (x_1-x_3,\bar{\tau}_1,\tau_3,\Omega_m)~,
\label{44}
\eea
where $L$ is the system size, the free Green's functions $g_\pm (x,\tau)$ are given by Eq.~(\ref{17}),
the interaction-induced factors $Q(x,\tau)$ are defined by Eq.~(\ref{Q}), and the factors
${\cal W}_\pm^{i,f}(x,\tau,\tau',\Omega_m)$ come from integration of the two
Green's functions attached to the current vertices over the external
coordinates and times:
\begin{eqnarray}
{\cal W}_{+}^{i} (x, \tau, \tau', \Omega_m) & =&  \displaystyle{ { {\rm sgn}  \Omega_m  \over
|\Omega_m |+v/l}}  \Big\{  \left(
e^{- i \Omega_m \tau} - e^{- i \Omega_m \tau'}  \right)
+ {v \over | \Omega_m | l} \left( 1 - e^{- |\Omega_m x| / v } \right)
\nonumber \\
& \times&   \left[e^{- i \Omega_m \tau}\theta(\Omega_m x )
-
e^{- i \Omega_m \tau'}
\theta ( -\Omega_m x ) \right]
\Big\},
\label{46}
\end{eqnarray}
\end{widetext}

\begin{eqnarray}
{\cal W}_-^{i,f}(x,\tau,\tau',\Omega_m)&=&-{\cal W}_+^{i,f}(-x,\tau,\tau',\Omega_m),
\nonumber \\
{\cal W}_+^f(x,\tau,\tau',\Omega_m)&=&{\cal W}_+^i(x,\tau,\tau',-\Omega_m).
\label{46-a}
\end{eqnarray}
In Eq.~(\ref{46}), we have included vertex corrections for the current
vertices, which arise from the anisotropy of impurity scattering [recall
that only backscattering off impurities (\ref{2}) is retained in the model].
The vertex corrections result in a replacement of the total scattering rate
$v/2l$ by the transport scattering rate $v/l$. Note also that Eq.~(\ref{44})
is written in terms of the contribution of the Cooperon with chiralities of
the current vertices as shown in Fig.~\ref{diagram}. The numerical coefficient
in Eq.~(\ref{44}) takes into account that $\Delta\sigma_{\rm WL}$ is a factor of
$2\times 2\times 2 =8$ larger than the contribution of the diagram in Fig~\ref{diagram}
[one of the factors of 2 comes from the spin, another from a summation over chiralities
of the current vertices and the third one from diagrams (b) and (c) in
Fig.~\ref{3diagrams}].

The approximation (\ref{19}) means that the terms in the exponent of
$Q(x,\tau)$ [Eqs.~(\ref{M}),(\ref{Q}), and(\ref{14})] that come from
$B_{++}(x,\tau)$ and $B_{--}(x,\tau)$ and are proportional to $\alpha_b$ are
neglected. As for the term that comes from $B_{+-}(x,\tau)$ and is
proportional to $\alpha_r$, it leads to a renormalization of the impurity
strength (see Ref.~\onlinecite{GMP} for details) but does not contribute to
the dephasing rate to first order in $\alpha$, similarly to the spinless
case. Therefore, in the calculation below we put both $\alpha_b$ and
$\alpha_r$ in $Q(x,\tau)$ equal to zero, while $l_0$ is replaced by $l$ which
is understood as the
renormalized mean free path. The factor $Q(x,\tau)$ in the limit of small
$\alpha$ is then written as
\bea
Q(x,\tau)\simeq \sqrt{{\sinh[\pi T(x/u+i\tau)]\sinh[\pi
T(x/u -i\tau)]\over\sinh [\pi T(x/v+i\tau)]\sinh [\pi T(x/v-i\tau)]}}~.
\nonumber\\
\label{45}
\eea
The second term in Eq.~(\ref{46}), proportional to $1/l$, can be omitted for
$l_\phi/l \ll 1$, so that one more approximation we make is to take the
factors (\ref{46}) in Eq.~(\ref{44}) at $x=0$:
\begin{equation}
{\cal W}_+^i(0,\tau,\tau',\Omega_m)={{\rm sgn} \Omega_m\over |\Omega_m|+v/l}
\left(e^{-i\Omega_m\tau}-e^{-i\Omega_m\tau'}\right)~.
\label{4.3a}
\end{equation}

It is convenient to introduce new variables
\bea
x_a &=& x_1 - x_3~, \nonumber \\
x_b &=& x_3 - x_2~, \nonumber \\
x_c &=& x_1 - x_2~,
\label{50}
\eea
and
\bea
\tau_a &=& \tau_1 - \bar{\tau}_3~, \nonumber \\
\tau_c &=& \tau_2 - \tau_1~, \nonumber \\
\tau_b &=& \tau_3 - \tau_2, \nonumber \\
\bar{\tau}_a &=& \bar{\tau}_1 - \tau_3~, \nonumber \\
\bar{\tau}_c &=& \bar{\tau}_2 - \bar{\tau}_1~, \nonumber \\
\bar{\tau}_b &=& \bar{\tau}_3 - \bar{\tau}_2~.
\label{49}
\eea
These satisfy the constraints $x_c = x_a+x_b$ and
$\tau_a+\tau_b+\tau_c+\bar{\tau}_a+\bar{\tau}_b+\bar{\tau}_c=0$. We thus
represent Eq.~(\ref{44}) as an integral over the variables
(\ref{50}),(\ref{49}) and insert, instead of $T/L$, in the integrand the factor
\be
T \sum_n e^{i \ve_n
(\tau_a+\tau_b+\tau_c+\bar{\tau}_a+\bar{\tau}_b+\bar{\tau}_c)}
\delta(x_c - x_a - x_b)
\label{51}
\ee
which contains summation over fermionic Matsubara frequencies. By extracting
the vertex functions (\ref{46}) at $x=0$, Eq.~(\ref{44}) in the limit $\alpha\ll 1$ is
rewritten in the new variables as
\begin{widetext}
\begin{eqnarray}
\frac{\Delta \sigma}{\sigma_{\rm D}} & \simeq & \lim_{\Omega\to 0} \,
\Big\{ \, -  \, {2  \pi  T  \over \Omega_m} \,\, {v^4 \over ( |\Omega_m |+ v/l)^2}
\,\,
\sum_{n} \int_{0}^{1/T} \!\!d \, \tau_a \, d \, \tau_b \, d \, \tau_c \,
d \, \bar{\tau}_a \, d \, \bar{\tau}_b \, d \, \bar{\tau}_c \,\,
\int d \, x_a \, d \, x_b \, d \, x_c \,
\nonumber \\
& \times &
G_{+} (x_a , \tau_a  ) \,\, G_{-} (x_a, \bar{\tau}_a ) \,\,
G_{+} (x_c, \tau_c) \,\, G_{-} (x_c, \bar{\tau}_c) \,\,
G_{+} (x_b , \tau_b ) \,\, G_{-} (x_b, \bar{\tau}_b ) \,
\nonumber \\
& \times &
C_{+} (x_a , \tau_a  ) \,\, C_{-} (x_a, \bar{\tau}_a ) \,\,
C_{+} (x_c, \tau_c) \,\, C_{-} (x_c, \bar{\tau}_c) \,\,
C_{+} (x_b , \tau_b ) \,\, C_{-} (x_b, \bar{\tau}_b ) \,
\nonumber \\
& \times & Q (x_a , \tau_b  + \tau_c ) \, Q (x_a , \bar{\tau}_b + \bar{\tau}_c)\,
Q (x_c , \tau_a  + \bar{\tau}_b) \, Q (x_c , \bar{\tau}_a + \tau_b ) \,
Q (x_b , \bar{\tau}_a + \bar{\tau}_c ) \, Q (x_b , \tau_a + \tau_c ) \,
\nonumber \\
& \times & \exp[\, i( \Omega_m + \ve_n )\, (\, \tau_a + \tau_b +  \tau_c \,) -
i \ve_n \, (\, \bar{\tau}_a + \bar{\tau}_b + \bar{\tau}_c \,) \,] \,\,
\delta \, (x_a + x_b - x_c) \, \Big\}_{i\Omega_m \to \Omega + i 0}  ,
\label{53}
\end{eqnarray}
where $G_\pm(x,\tau)$ is given by Eq.~(\ref{19}), $Q(x,\tau)$ by
Eq.~(\ref{45}), and we introduce
\be
C_\pm(x,\tau)=\sqrt{{\sinh [\pi T(x/v\mp i\tau)]\over
\sinh [\pi T(x/u\mp i\tau)]}}~,
\label{48}
\ee
such that $g_{\pm} Q^{-1} \simeq G_{\pm} C_{\pm}$. When deriving Eq.~(\ref{53}) we have used
the approximation (\ref{4.3a})
\begin{eqnarray}
{\cal W}_{+}^{i} (0, \tau_1 , \bar{\tau}_3 ,  \Omega_{m} )
{\cal W}_{-}^{f} (0, \bar{\tau}_1 , \tau_3 , \Omega_{m}) &\to&
\displaystyle{
{1  \over
( |\Omega_m |+ v/l )^2 }
} \,
\Big[ -
e^{ i \Omega_m ( \tau_1 - \bar{\tau}_1)} -
e^{ i \Omega_m (\bar{\tau}_3 - \tau_3)}
\nonumber \\
&+&
e^{ i \Omega_m ( \tau_1 - \tau_3)} +
e^{ i \Omega_m (\bar{\tau}_3 - \bar{\tau}_1)}
\Big] \to - \displaystyle{
{2 \,  e^{ i \Omega_m (\bar{\tau}_3 - \tau_3)} \over
( |\Omega_m |+ v/l)^2 }}.
\label{48-a}
\end{eqnarray}
\end{widetext}
The ``diagonal'' terms $\exp[i\Omega_m(\bar{\tau}_1-\tau_1)]$ and
$\exp[i\Omega_m(\tau_3-\bar{\tau}_3)]$ yield identical contributions [which is accounted
for by the factor of 2 in Eq.~(\ref{53})]. The cross-terms are neglected, since, after the integration over
times (see Secs.~\ref{IVb} and \ref{V} below), they produce the products of Green's functions in the $(x,\ve)$
representation of the type
\begin{equation}
 G^{r(a)}_{+}(x,\ve_n)G^{r(a)}_{-}(x,\ve_k)=0
\label{rr}
\end{equation}
[see Eq.~(\ref{27})]. This corresponds to retaining only those Cooperon diagrams that contain an equal number
of the retarded and advanced Green's functions, even with e-e interaction included.

We are thus left with the functions $G_\pm (x,\tau)$, $Q(x,\tau)$, and $C_\pm
(x,\tau)$ that are all given by various combinations of square-root factors
$\sqrt{\sinh [\pi T(x/u\pm i\tau)]}$ and $\sqrt{\sinh [\pi T(x/v\pm
i\tau)]}$. Note that $G_\pm (x,\tau)$ and $C_\pm (x,\tau)$ enter Eq.~(\ref{53})
only in the combination $G_\pm C_\pm$ with the same arguments. Equation
(\ref{53}) will be analyzed in the next two subsections.

\subsection{Regular impurity configurations: Weak localization}
\label{IVb}

In Eq.~(\ref{53}), we transform the integration contours for each of the time
variables similarly to Fig.~\ref{complex}. As a result, we obtain integrals
along square-root branch cuts in the vertical direction, each of which
connects two points whose coordinates can be written as $\tau+ix/u$ and
$\tau+ix/v$ with different $\tau$ and $x$. For example, let us assume, here
and throughout the paper below, that in Eq.~(\ref{53}) all
\be
x_a,x_b,x_c>0
\ee
(the region of integration $x_a<0$ and $x_b,x_c>0$ gives the same contribution
to $\Delta\sigma_{\rm WL}$). Then, starting with the integration over $\tau_a$ and
closing the contour upwards, we represent the integral along the real axis of
$\tau_a$ as a sum of three integrals around vertical cuts: between $ix_a/u$
and $ix_a/v$, between $-\bar{\tau}_b+ix_c/u$ and $-\bar{\tau}_b+ix_c/v$, and
between $-\bar{\tau}_c+ix_b/u$ and $-\bar{\tau}_c+ix_b/v$. The first cut comes
from the Green's function $G_+(x_a,\tau_a)$, whereas the last two from the
factors $Q(x_b,\tau_a+\tau_c)$ and $Q(x_c,\tau_a+\bar{\tau}_b)$,
respectively. Since $G_+(x,\tau)$ as a function of $x$ for a given $\tau$
falls of on the scale of $l_T$, while $Q(x,\tau)$ on the scale of $l_{ee}$,
one sees that in the limit of small $\alpha$ the main contribution to
$\Delta\sigma_{\rm WL}$ comes from the branch cuts that are associated with the
Green's functions. The cuts related to the factors $Q(x,\tau)$ can be
neglected.

The selection of singularities at $\alpha\ll 1$ is closely analogous to that
in the spinless case, see Appendix F of Ref.~\onlinecite{GMP}. For spinless
electrons, the main contribution to $\Delta\sigma_{\rm WL}$ stems from
singularities (``nearly poles" for weak interaction) of the single-particle
Green's functions at the classical trajectory of an electron moving with the
velocity $u$. Other close-to-pole singularities (those in the factors
$Q$ and $C$), which are related to ``nonclassical
trajectories'', yield subleading corrections small in the parameter
$l_T/l_\phi\ll 1$. The spinful problem is very much similar in this respect.
The main difference is that the dominant singularities are now pairs
of close square-root branching points rather than the poles. The transformation of the
poles into the branch
cuts, induced by the SCS, can be viewed as a ``smearing" of the classical
trajectories: all velocities between $v$ and $u$ become accessible.

Let us first consider the contribution to Eq.~(\ref{53}) from typical impurity
configuration for which the characteristic scale
of
\be
x_a\sim x_b\sim x_c
\label{typ}
\ee
is of the order of $l_\phi\gg l_T$. We can then expand all $\sinh$'s in the
$C$ and $Q$ factors as
\be
\sinh(\pi Ty)\simeq {1\over 2}e^{\pi T|y|}\,{\rm sgn}\,y~,
\qquad |y|\gg 1~.
\label{56}
\ee
It is immediately seen that using Eq.~(\ref{56}) reduces the product of six
$C$ factors to a simple exponential:
\be
{\cal C}\simeq\exp (2x_c/l_{ee})~.
\label{56a}
\ee
Similarly, the product of six $Q$ factors
\begin{widetext}
\bea
{\cal Q}&=&
Q\left[x_a,i\left(t_c+t_b+{2x_b+x_a\over u}\right)\right]
Q\left[x_a,i\left(\bar{t}_c+\bar{t}_b+{2x_b+x_a\over u}\right)\right]
Q\left[x_c,i\left(t_a-\bar{t}_b+{x_a-x_b\over u}\right)\right]
\nonumber\\
&\times&
Q\left[x_c,i\left(t_b -\bar{t}_a + {x_b-x_a\over u}\right)\right]
Q\left[x_b,i\left(t_a+t_c+{2x_a+x_b\over u}\right)\right]
Q\left[x_b,i\left(\bar{t}_a+\bar{t}_c+{2x_a+x_b\over u}\right)\right]
\label{57}
\eea
is represented as
\be
{\cal Q}\simeq\exp (\pi Tq/2)~,
\label{Qq}
\ee
where
\bea
q&=&
|2x_b/u+t_c+t_b|-|2x_b/u+t_c+t_b-2x_a/u_-|+
|2x_c/u+t_c+t_b|-|2x_c/u+t_c+t_b+2x_a/u_-|
\nonumber\\
&+&
|2x_b/u+\bar{t}_c+\bar{t}_b|-|2x_b/u+\bar{t}_c+\bar{t}_b-2x_a/u_-|+
|2x_c/u+\bar{t}_c+\bar{t}_b|-|2x_c/u+\bar{t}_c+\bar{t}_b+2x_a/u_-|
\nonumber\\
&+&
|2x_b/u-t_a+\bar{t}_b|-|2x_b/u-t_a+\bar{t}_b+2x_c/u_-|+
|2x_a/u+t_a-\bar{t}_b|-|2x_a/u+t_a-\bar{t}_b+2x_c/u_-|
\nonumber\\
&+&
|2x_a/u+\bar{t}_a-t_b|-|2x_a/u+\bar{t}_a-t_b+2x_c/u_-|+
|2x_b/u-\bar{t}_a+t_b|-|2x_b/u-\bar{t}_a+t_b+2x_c/u_-|
\nonumber\\
&+&
|2x_a/u+t_a+t_c|-|2x_a/u+t_a+t_c-2x_b/u_-|+
|2x_c/u+t_a+t_c|-|2x_c/u+t_a+t_c+2x_b/u_-|
\nonumber\\
&+&
|2x_a/u+\bar{t}_a+\bar{t}_c|-|2x_a/u+\bar{t}_a+\bar{t}_c-2x_b/u_-|+
|2x_c/u+\bar{t}_a+\bar{t}_c|-|2x_c/u+\bar{t}_a+\bar{t}_c+2x_b/u_-|~.
\label{58}
\eea
In Eqs.~(\ref{57}) and (\ref{58}) we have shifted the time variables according to
\begin{eqnarray}
i \tau_j &=& - t_j - x_j / u , \nonumber \\
i \bar{\tau}_j &=&  \bar{t}_j  +  x_j / u ,
\label{54}
\end{eqnarray}
with $j=a,b,c$. The shifted variables $t_j$ and $\bar{t}_j$ in
Eq.~(\ref{54}) are real on the branch cuts corresponding to the
Green's functions in Eq.~(\ref{53}) and change along these cuts
from 0 to $2x_j/u_-\sim\alpha x_j/u$, where $u_-$ is given by
Eq.~(\ref{24}). Inspecting Eq.~(\ref{58}), we observe that for
$x_a>2x_c u/u_-$ (or similarly for $x_b$) all the moduli can in
fact be omitted on the Green's function branch cuts, which yields
\be
{\cal Q} \simeq \exp (- 4 \pi T x_c/u_-)=\exp (-2x_c/l_{ee})
\label{59}
\ee
(the opposite case $x_{a,b}<2\alpha x_c$ is
addressed in Sec.~\ref{V}). As a result, the factors $\cal C$ and
$\cal Q$ compensate each other: \be {\cal QC}\simeq 1~. \ee The
integrand of Eq.~(\ref{53}) thus reduces to a product of the
single-particle Green's functions in the $(x,\ve)$ representation
[Eqs.~(\ref{27}), (\ref{29})]:
\begin{eqnarray}
\frac{\Delta \sigma_{\rm WL}}{\sigma_{\rm D}} & = &\lim_{\Omega\to 0} \Big\lbrace
\, - \, {2  \pi  T  \over \Omega_m} \,\, {v^4 \over (|\Omega_m |+ v/l )^2}
\,\,
\sum_{n} \,  \int_0^\infty  d  x_a  \,  d  x_b \,  d  x_c  \, \delta \, (x_a +
x_b - x_c)
\nonumber \\
& \times &
G_{+}^{r} (x_a , i\ve_n + i\Omega_m)
G_{+}^{r} (x_c, i\ve_n + i\Omega_m)
G_{+}^{r} (x_b , i\ve_n + i\Omega_m)
G_{-}^{a} (x_a, i\ve_n)
G_{-}^{a} (x_c, i\ve_n)
G_{-}^{a} (x_b, i\ve_n) \Big\rbrace_{i\Omega_m \to \Omega+i0}~, \nonumber \\
\label{58-1}
\end{eqnarray}
where we have taken into account that only terms with $\ve_n<0, \
\ve_n+\Omega_m>0$ survive, in view of Eq.~(\ref{rr}), which
follows from Eq.~(\ref{27}). Performing the analytical continuation to real
frequencies $\Omega_m\to\Omega+i0$, we get in the dc limit $\Omega\to 0$:
\be
{\Delta \sigma_{\rm WL} \over
\sigma_{\rm D}} = -A_{\rm WL} \Big( {l_{ee} \over l} \Big)^2\ ,
\label{60}
\ee
where the numerical factor $A_{\rm WL}$ is defined by
\be
\label{60-1}
A_{\rm WL}= \pi \int_{-\infty}^{\infty} \, {dz \over {\rm cosh}^2 \pi z} \,
\int_{0}^{\infty}  dx \, \int_{0}^{\infty}  dy \, {\cal R} (x,z) \, {\cal R} (y,z) \,
{\cal R} ( x + y + xy,z),
\ee
with
\be
\label{60-2}
{\cal R} (x,z) = \,
_{2}{\rm F}_{ 1} \left(\, 1/2 + i z , \, 1/2, \, 1; \, -x  \,\right) \,
_{2}{\rm F}_{ 1} \left(\, 1/2 - i z , \, 1/2, \, 1; \, -x  \,\right).
\ee
We have estimated $A_{\rm WL}\simeq 0.13$ by taking the integral (\ref{60-1})
numerically.
\end{widetext}

The small factor $(l_{ee}/l)^2$ in Eq.~(\ref{60}) is due to the
exponential decay $\exp (-2x_c/l_{ee})$ of the product of six Green's
functions in the integrand of Eq.~(\ref{58-1}) on the scale of $l_{ee}$. One
sees that the dephasing factor that suppresses the interference term in the
conductivity behaves as $\exp (-L_C/l_\phi)$, where $L_C=x_a+x_b+x_c=2x_c$ is the total
length of the Cooperon loop and the WL dephasing length $l_\phi$ reads
\be
l_\phi=l_{ee}~.
\label{62}
\ee
Schematically, Eq.~(\ref{58-1}) can be estimated as
\bea
\label{power}
{\Delta\sigma_{\rm WL}\over\sigma_{\rm D}}\sim
&-&\int_0^\infty \frac{dx_a}{l} \exp (-2x_a/l_{ee})
\nonumber\\
&\times&\int_0^\infty\frac{dx_b}{l} \exp (-2x_b/l_{ee}) \sim - \Big({l_{ee} \over l}\Big)^2~,
\nonumber \\
\eea
so that for typical realizations of disorder with $x_a\sim x_b\sim
x_c$ each of the distances is of the order of $l_{ee}$, see Fig.~\ref{scales}a.
For comparison, in the spinless case,\cite{GMPlett,GMP}
the relevant distances obey $x_a x_b \sim l l_{ee}$, which yields
\bea
\label{powersl}
{\Delta\sigma_{\rm WL}^{sl} \over\sigma_{\rm D}}
&\sim& - \int_0^l \frac{dx_a}{l} \int_0^l\frac{dx_b}{l}\, \exp (-x_a x_b / l l_{ee})
\nonumber\\
&\sim & - {l_{ee} \over l}  \, \ln {l \over l_{ee}} \sim -
\left({l_{\phi} \over l}\right)^2 \, \ln {l \over l_{\phi}} ~.
\nonumber \\
\eea
It follows that $l_\phi$ for spinful electrons is much shorter than the
dephasing length for spinless electrons (recall that for the latter, $l_\phi$
diverges\cite{GMPlett,GMP} in the limit of vanishing disorder). In fact, in the
spinful case $l_\phi$ is equal to the single-particle (electron) decay length
$l_{ee}$, in contrast to the spinless case, where $l_\phi\gg
l_{ee}$.

\section{Anomalous Impurity Configurations: Memory Effects}
\label{V}
\setcounter{equation}{0}

\begin{figure}[ht]
\centerline{
\includegraphics[width=7.0cm]{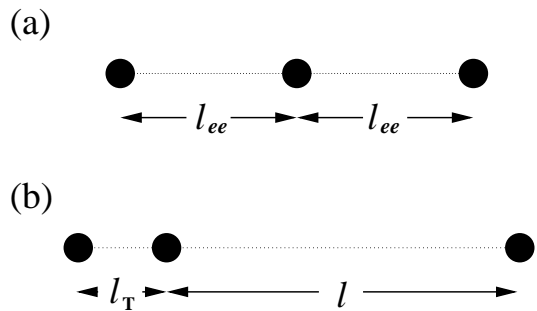}
}
\caption{Three-impurity configurations that give the main contribution to the
(a) quantum (WL) and (b) classical (ME) corrections to the
conductivity. The characteristic distances between the impurities are shown.
}
\label{scales}
\end{figure}

The WL contribution to the conductivity, calculated in the preceding section,
is associated with scattering on compact three-impurity configurations which
are ``regular" in the sense that the characteristic distances between all
three impurities are the same. Below, we will see that ``anomalous" (strongly asymmetric)
configurations in which two of the impurities are anomalously close to each
other, i.e., $x_a\ll x_b$ or $x_b\ll x_a$ (see Fig.~\ref{scales}b), give rise
to a conductivity correction which is larger than $\Delta\sigma_{\rm WL}$ if $T$
is sufficiently high. As mentioned already in Sec.~\ref{Intro}, the relevance
of the asymmetric configurations is related to the {\it classical}
ME\cite{zero-omega} in electron kinetics, in contrast to the {\it quantum}
interference of scattered waves that yields $\Delta\sigma_{\rm WL}$.

%%%%%%%%%%%%%%%%%%%%%%%%%%%%%%%%%%%%%%%%%%%%%%%%%%%%%%%

%%%%%%%%%%%%%%%%%%%%%%%%%%%%%%%%%%%%%%%%%%%%%%%%%%%%%%%%%

\subsection{Qualitative consideration: Identifying scales and parameters}
\label{Va}

To demonstrate the peculiarity of the asymmetric impurity configurations, it
is instructive to consider first the limit of two scattering events occurring
at the same point, by setting
\be
x_1 = x_3
\label{103}
\ee
in Eq.~(\ref{44}).  As discussed at the beginning of Sec.~\ref{IV}B, for
typical impurity configurations (\ref{typ}) the main
contribution to the correction (\ref{44}) comes from ``smeared'' classical
trajectories, meaning that all trajectories with velocities between $v$ and
$u$ contribute to $\Delta\sigma_{\rm WL}$. The case (\ref{103}) is, however,
special in that only one velocity remains and that is the velocity of
noninteracting electrons $v$. Indeed, at $x_1=x_3$ the four $Q$ factors that
depend on $x_1-x_3$ drop out of the integration over times---since
$Q(0,\tau)$ [Eq.~(\ref{45})] does not depend on $\tau$.
After this, the integrals in Eq.~(\ref{44}) over $\tau_1$ and $\bar{\tau}_1$
are dominated by the poles of the noninteracting Green's functions $g_{+} (0, \tau_1 - \bar{\tau}_3)$
and $g_{-} (0, \bar{\tau}_1 - \tau_3)$, which yields
\be
\tau_1 = {\bar \tau}_3, \quad {\bar \tau}_1 = \tau_3,
\label{103a}
\ee
and then all the remaining factors $Q$ compensate each other. We thus end up with a
product of six bare Green's functions $g_{\pm}$ multiplied by the factors
${\cal W}_{\pm}$, which constitutes the {\it noninteracting} limit of the problem.\cite{footnon}

In fact, the compensation of the dephasing factors is
evident even before the averaging over the fluctuating fields $\theta_\pm
(x,\tau)$ (Fig.~\ref{diagram}). Indeed, the factors $\exp \{\pm i
[\theta_+(x,\tau)-\theta_-(x,\tau)]\}$, dressing two impurity vertices,
cancel each other when taken at the same space-time point [Eqs.~(\ref{103}) and (\ref{103a})].
As a result, the third impurity
located at $x_2$ becomes decoupled with respect to e-e interaction
from the two-impurity complex at $x_1 = x_3$.

%%%%%%%%%%%%%%%%%%%%%%%%%%%%%%%%%%%%%%%%%%%%%%%%%%%%%%%%%%%%%%%%%%%%%%%%%

\begin{figure}[ht]
\centerline{
\includegraphics[width=8cm]{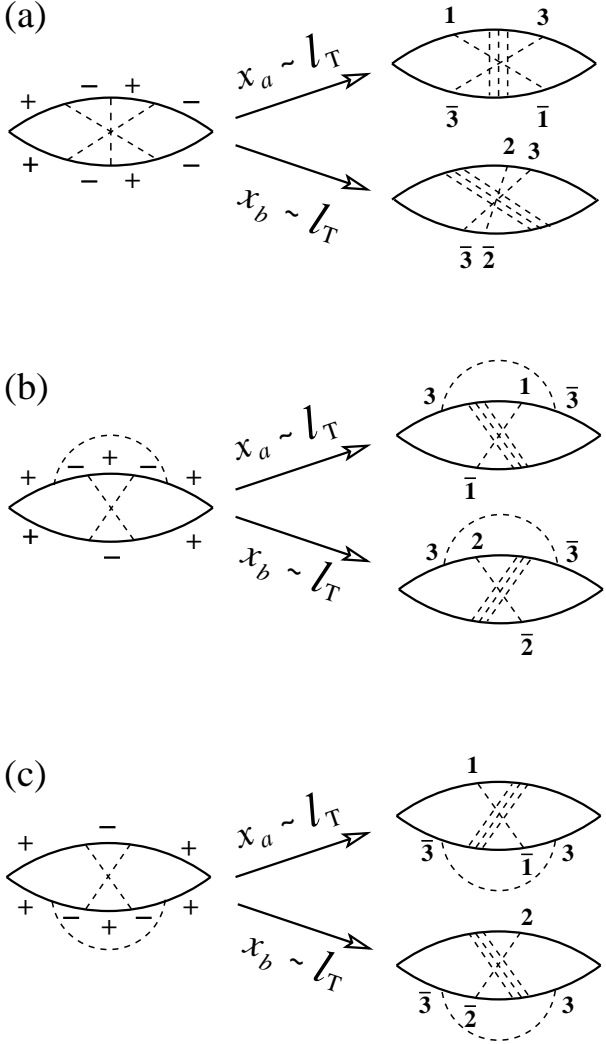}
}
\caption{
Each diagram for $\Delta\sigma$ in Fig.~\ref{3diagrams} has two ME contributions, coming from
$x_a\sim l_T$ ($\bar{3}\to 1$, $\bar{1}\to 3$) and $x_b\sim l_T$
($2\to 3$, $\bar{2}\to \bar{3}$). The remote-impurity line should be replaced
by the full diffuson ladder.}
\label{3diagramsME}
\end{figure}

%%%%%%%%%%%%%%%%%%%%%%%%%%%%%%%%%%%%%%%%%%%%%%%%%%%%%%%%%%%%%%%%%%%%%%%%%%%%%%%

In the cyclic variables (\ref{50}), Eq.~(\ref{103}) means $x_a =0$.  Since
interaction completely drops out of the problem at $x_a=0$, the integration
over the remaining spatial coordinate $x_b = x_c$ in Eq.~(\ref{53}) is not cut
off by dephasing, in contrast to the regular impurity configurations, for
which it is restricted to $x_c\alt l_{ee}\ll l$. In this situation, we have to
take into account a disorder-induced damping of the single-particle Green's
functions $g_{\pm}(x,\tau)$, which results in an additional factor $\exp
(-|x|/4l)$ attached to each $g_{\pm}(x,\tau)$. On the Cooperon loop, these
combine to give the overall factor $\exp (-x_c /l)$ in the integrand of
Eq.~(\ref{44}), so that $x_b=x_c$ is then limited by the mean free path, see
Fig.~\ref{scales}b.

Since at $x_a=0$ the characteristic distance to the remote third impurity in
the three-impurity Cooperon diagram happens to be in the crossover region
between the ballistic and diffusive motion, we should extend the single
scattering on the third impurity to an infinite sequence of scatterings on
other impurities,\cite{footnote-ladder} i.e., to a diffuson ladder, as shown
in Fig.~\ref{3diagramsME}. One sees that the diagram takes the form
characteristic of a quasiclassical ME:\cite{zero-omega} an electron is
scattered at $x_1\simeq x_3$, then moves around diffusively, and returns to
$x_1\simeq x_3$ where it is scattered once again. Clearly, this is a
non-Markovian process which is beyond the conventional Boltzmann
description. However, the non-Boltzmann type of kinetics associated with the
return processes is classical in origin: as discussed above, when the points
$x_1$ and $x_3$ are sufficiently close to each other, dephasing becomes
irrelevant. We will first analyze the simplest three-impurity diagram at
finite but small $x_a$ and include the diffusive returns (which will only
renormalize the numerical prefactor) later in the end of this section.

It is instructive to begin with a semi-quantitative analysis which will give
correctly the parametric dependence of the result but not the numerical
coefficient. To this end, we replace all hyperbolic sines in both the Green's
functions $G_{\pm}$ and the $Q$ and $C_\pm$ factors in Eq.~(\ref{53}) by their
exponential asymptotics, Eq.~(\ref{56}). This approximation is parametrically
correct, since all integrals in Eq.~(\ref{56}) are determined by the regions
of integration in which the arguments of the hyperbolic sines are of the order
of or larger than 1. Within the ``exponential'' approximation, the product
${\cal C}$ of six factors $C_{\pm}$ is given by Eq.~(\ref{56a}), while the
product ${\cal G}$ of six Green's functions in Eq.~(\ref{56}) becomes
\be
{\cal G}\propto\exp (-2x_c/l_{ee})~,
\label{56b}
\ee
so that the exponential factors in $\cal G$ and $\cal C$ cancel each other,
similarly to the regular impurity configurations in Sec.~\ref{IVb}.  What is
different, however, is that when calculating the factor ${\cal Q}$
[Eqs.~(\ref{57})--(\ref{58})] one can no longer omit the moduli in
Eq.~(\ref{58}) if
\be
x_a < 2\alpha x_c
\label{asym}
\ee
[or $x_b<2\alpha x_c$---we will proceed with the estimate for the case of
Eq.~(\ref{asym})].  For the strongly asymmetric configurations (\ref{asym}),
Eq.~(\ref{58}) is rewritten as
\begin{widetext}
\begin{eqnarray}
q & \simeq & -8 x_c / u_{-} + \big| 2 x_a / u - t_b \big|
- \big| 2 x_a / u - t_b + 2 x_c / u_{-} \big| +
\big| 2 x_a / u - \bar{t}_b \big| - \big| 2 x_a / u
- \bar{t}_b + 2  x_c / u_{-} \big|
\nonumber \\
& & +
\big| 2 x_a / u + t_c \big| - \big| 2 x_a / u + t_c
- 2  x_c / u_{-} \big| +
\big| 2 x_a / u + \bar{t}_c \big| - \big| 2 x_a / u + \bar{t}_c
- 2 x_c / u_{-} \big|~,
\label{600}
\end{eqnarray}
where we have neglected terms of second order in $\alpha$. In particular, the
integration over $t_a$ and $\bar{t}_a$ goes along very short branch cuts of
length $\propto\alpha x_a \sim \alpha^2 x_c$. To our accuracy, the cuts reduce to
poles. For this reason, we have set $t_a=\bar{t}_a=0$ in Eq.~(\ref{600}).

The integrals over remaining time variables in Eq.~(\ref{53}) decouple from
each other and can be easily calculated (note that the integrals over $t_b$
and $t_c$ are identical---in the dc limit---to those over $\bar{t}_b$ and $\bar{t}_c$,
respectively). Let us denote $\Delta\sigma_{\rm asym}$ the contribution to the
conductivity that comes from the strongly asymmetric impurity configurations
with $|x_{a,b}|<2\alpha x_c$. Carrying out the analytical continuation for both
fermionic and bosonic frequencies and taking the dc limit, we get, within the
exponential approximation:
\begin{eqnarray}
\frac{\Delta \sigma_{\rm asym}}{\sigma_{\rm D}}&\sim&
-{1 \over l^2}\int_{0}^{\infty}\!{d\ve \,dx_a\,dx_c\over T \cosh^2(\ve/2T)}
\exp\left(-{x_c\over l}\right)
\exp\left(-{2x_c\over l_{ee}}\right)
\nonumber\\
&\times&\left[\,\theta (x_a-\alpha x_c)\theta (2\alpha x_c-x_a)\,
\Big| f_1\left( {x_a\over 2 l_T}, {x_c\over 2 l_{ee}},
{\ve\over \pi T} \right)    \Big|^4 \right.
\nonumber \\
&+&\left.\theta ( \alpha x_c-x_a)\,\Big|
f_2\left( {x_a \over 2 l_T}, {x_c \over 2 l_{ee}}, {\ve \over \pi T}\right)
\Big|^2\,\Big| f_3\left( {x_a\over 2 l_T},{x_c\over 2 l_{ee}}~,
{\ve\over\pi T}\right)\Big|^2\,\right]~,
\label{100}
\end{eqnarray}
where the functions $f_{1,2,3}$ are defined as follows
\begin{eqnarray}
f_1 (x,y,z)& = &  {\exp ( iz y ) - 1 \over iz },
\nonumber \\
f_2 (x,y,z)& = & \exp [ iz ( y - x ) ] {\exp ( iz x )  - 1 \over iz }
- i \exp ( x-y ) {\exp [( iz + 1 ) ( y -x )]  - 1 \over ( iz + 1  )},
\nonumber \\
f_3 (x,y,z)& = &  {\exp ( iz x )  - 1 \over iz }
+ i \exp ( -  x  ) {\exp [( iz + 1 ) y ]
-  \exp [( iz + 1 ) x ] \over ( iz + 1  )}~.
\label{101}
\end{eqnarray}
In Eq.~(\ref{100}), the $\theta$-functions split the domain of integration
over $x_a$ into two, $\alpha x_c<x_a<2\alpha x_c$ and
\be
0<x_a<\alpha x_c~.
\label{ME}
\ee
In the first interval of $x_a$, the integration over $x_c$ is limited by the
factor $\exp (- 2 x_c / l_{ee})$ in the first line of Eq.~(\ref{100}), which
yields a contribution to $\Delta\sigma_{\rm asym}/\sigma_{\rm D}$ of the order of $-l_{ee}l_T/l^2$.
%\be
%-\left[\int_0^\infty\! {dx_c\over l}\exp \left(-{2x_c\over l_{ee}}\right)
%\right] \,
%\left[\int_0^\infty\! {dx_a\over l}\exp \left(-{2x_a\over l_T}\right)\right]
%\sim - {l_{ee} \over l}\, {l_T \over l}~.
%\ee
This is much smaller than the contribution of the regular impurity
configurations [Eq.~(\ref{59})] and can be neglected. However, the situation is
qualitatively different for the interval (\ref{ME}). Indeed, for $1,x\ll y$,
the functions $|f_2(x,y,z)|^2$ and $|f_3(x,y,z)|^2$ take the form
\begin{eqnarray}
|f_2(x,y,z)|^2&\simeq& \Big| {\exp ( i z x) - 1 \over  z} +  {1 \over 1 + i z} \Big|^2 ~,
\label{102-1} \\
|f_3(x,y,z)|^2&\simeq&{\exp [2(y-x)]\over (z^2+1)}~.
\label{102}
\end{eqnarray}
As a result, the factor $\exp (-2 x_c/l_{ee})$ from the first line of
Eq.~(\ref{100}) is canceled by the factor $\exp(2y)$ from
Eq.~(\ref{102}). The remaining integral over $x_c$ is no longer restricted to
$x_c\alt l_{ee}$ but rather extends to $x_c\sim l$, in agreement with our
qualitative consideration of the ME effect for $x_a=0$ at the beginning of
this section. The ME contribution $\Delta\sigma_{\rm ME}$ to the conductivity 
thus arises from the impurity configurations obeying Eq.~(\ref{ME}) and is 
given by the second term in Eq.~(\ref{100}). It can be estimated as
\be
{\Delta\sigma_{\rm ME}\over\sigma_{\rm D}}\sim
-\int_0^\infty\! {dx_c\over l}\exp \left(-{x_c\over l}\right)
\int_0^{\alpha x_c}\! {dx_a\over l}\exp \left(-{2x_a\over l_T}\right)
\sim -{l_T \over l}
\label{104}
\ee
and becomes much larger than the WL contribution [Eq.~(\ref{59})] if $T$ is
sufficiently high, namely if
\be
T\gg v/\alpha^2l~.
\ee

%%%%%%%%%%%%%%%%%%%%%%%%%%%%%%%%%%%%%%%%%%%%%%%%%%%%%%%%%%%%%%%%%%%%%%%%%
%%%%%%%%%%%%%%%%%%%%%%%%%%%%%%%%%%%%%%%%%%%%%%%%%%%%%%%%%%%%%%%%%%%%%%%%%

\begin{figure}[ht]
\centerline{
\includegraphics[width=14cm]{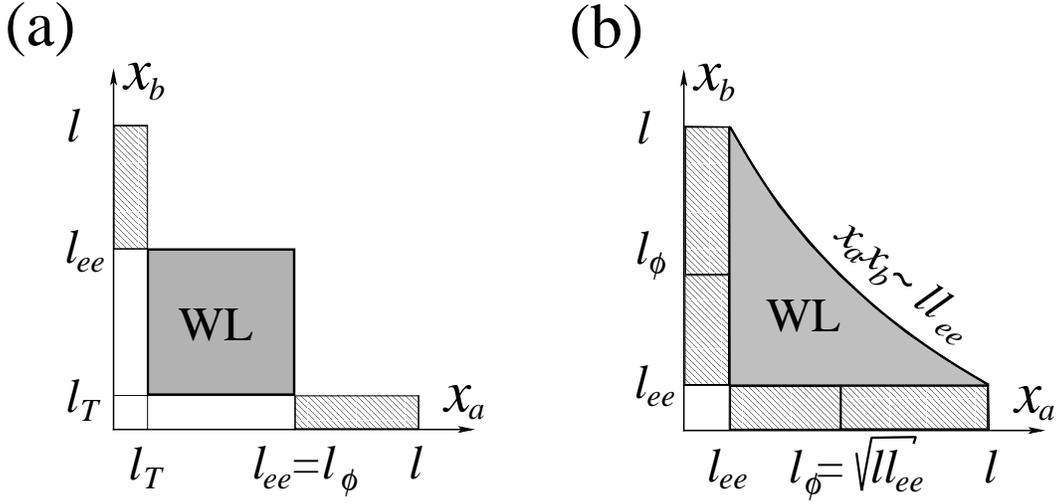}
}
\caption{
Domains of the distances $x_a,x_b$ that govern the WL (gray-shaded area) and ME (dashed area)
contributions to the conductivity for the spinful (a) and spinless (b) models.
}
\label{xaxb}
\end{figure}

%%%%%%%%%%%%%%%%%%%%%%%%%%%%%%%%%%%%%%%%%%%%%%%%%%%%%%%%%%%%%%%%%%%%%%%%%%%%%%

The characteristic regions of $x_a,x_b$ that control the quantum (WL) and
classical (ME) corrections to the conductivity are shown in Fig.~\ref{xaxb}a.
For comparison, an analogous plot for the spinless problem is presented in
Fig.~\ref{xaxb}b. One sees that in the spinful case the ``quantum'' and
``classical'' domains are parametrically separated from each other. As a
result, their areas $l_{ee}^2\propto T^{-2}$ and $l l_T\propto T^{-1}$,
respectively, can ``compete" with each other. On the contrary, in the spinless
case, the ME strips of area $\sim l l_{ee}$ are directly adjacent to the WL
domain, whose area $\sim l l_{ee} \ln(l/l_{ee})$ is (logarithmically) larger.
It follows that for spinless electrons the ME correction
$\Delta\sigma^{sl}_{\rm ME}$ gives only a subleading contribution as compared to
the WL correction $\Delta\sigma^{sl}_{\rm WL}$ [Eq.~(\ref{powersl})]:
\begin{equation}
 \left|{\Delta\sigma^{sl}_{\rm ME}\over \sigma_{\rm D}}\right|
\sim {l_{ee}\over l} \ll
\left|{\Delta\sigma^{sl}_{\rm WL}\over \sigma_{\rm D}}\right|
%\sim 
%\left(\frac{l_\phi}{l}\right)^2 \ln\left(\frac{l}{l_\phi}\right)
\sim {l_{ee}\over l} \ln\left(\frac{l}{l_{ee}}\right)
\end{equation}
[recall that $l_\phi\sim (l l_{ee})^{1/2}$ and $l_{ee}\sim \alpha^{-2} l_T$ for
spinless electrons].

\subsection{Rigorous calculation of $\Delta\sigma_{\rm ME}$}
\label{Vb}

In the preceding subsection, we have adopted the ``exponential approximation"
by replacing all the hyperbolic sines in the correlation functions by their
asymptotics (\ref{56}). This allowed us to identify the relevant scales and
obtain the parametric dependence of the result. Here, we calculating the
integrals over the temporal variables $t_b$, ${\bar t}_b$, $t_c$, and ${\bar
t}_c$ more accurately and find the numerical prefactor in Eq.~(\ref{104}).
The estimate in Sec.~\ref{Va} teaches us that the dominant contribution to
$\Delta\sigma_{\rm ME}$ comes from $x_a\sim l_T$ and $x_b\simeq x_c\sim l$, while
the integrals over $t_{b,c}$ and ${\bar t}_{b,c}$ are determined by the upper
limit of integration, i.e., by the time scale $\alpha x_c/v\sim \alpha
l/v$. In the WL regime, this time scale is much larger than $1/T$.

Similarly to the case of regular impurity configurations, we observe that the
product of all the $C$ factors can always be approximated by Eq.~(\ref{56a}),
whereas the hyperbolic sines should be retained in the Green's functions
$G_{\pm}$. As for the factors $Q$, in contrast to the regular configurations,
not all of them can be replaced by their asymptotics. Specifically, those
hyperbolic sines in the $Q$ factors that correspond to the moduli in
Eq.~(\ref{600}) (i.e., all terms in $q$ except for the first one) should
retain their form and not be replaced by the exponentials. Importantly, within
this ``partially exponential'' approximation, all integrals over times are
still decoupled. Thus, we replace the exponential factor depending on $t_c$ in
Sec.~\ref{Va} according to
\begin{eqnarray}
&&  \Big[ \,
\theta \big( x_a / l_T + \pi T t_c - x_c / l_{ee} \big) - i \theta
\big( - x_a / l_T - \pi T t_c + x_c / l_{ee} \big) \, \Big] \nonumber \\
&& \times \,\, \exp \left[{1 \over 2} \left(
-x_c / l_{ee} + | x_a / l_T + \pi T t_c | - | x_a / l_T + \pi T t_c
- x_c / l_{ee} |\right) - |\ve_n +\Omega_m | \, t_c \right]
\nonumber \\
&&\rightarrow
{ 2 \exp (- |\ve_n +\Omega_m | \, t_c ) \over \sinh^{1/2}
( \pi T t_c )  \sinh^{1/2} (x_c / l_{ee} - \pi T t_c )    }
\, {\sinh^{1/2} (x_a / l_T + \pi T t_c ) \over \sinh^{1/2} (x_a / l_T
+ \pi T t_c - x_c / l_{ee} )}\nonumber \\
&&\rightarrow
{ 2 \exp ( x_a/2l_T)\exp (- |\ve_n +\Omega_m | \, t_c)
\over  \sinh^{1/2} (x_c / l_{ee} - \pi T t_c )
\sinh^{1/2} (x_a / l_T + \pi T t_c - x_c / l_{ee} ) }
\label{105}
\end{eqnarray}
and make a similar replacement for ${\bar t}_c$ (with $\ve_n +\Omega_m \to
\ve_n$). The factors depending on $t_b$ are modified as follows:
\begin{eqnarray}
&&i \,\exp \left[{1 \over 2} \left(
-x_c / l_{ee} + | x_a / l_T - \pi T t_b | - | x_a / l_T - \pi T t_b
+ x_c / l_{ee} |\right) - |\ve_n +\Omega_m | \, t_b \right]
\nonumber \\
&&\rightarrow
{ 2 \exp \left(- |\ve_n +\Omega_m | \, t_b \right) \over \sinh^{1/2} (
\pi T t_b )  \sinh^{1/2} (x_c / l_{ee} - \pi T t_b )    }
\, {\sinh^{1/2} (x_a / l_T - \pi T t_b) \over \sinh^{1/2} (x_a / l_T -
\pi T t_b + x_c / l_{ee} )}
\nonumber \\
&&\rightarrow
{ 2 \exp\left(-x_a/2l_T\right) \exp \left(- |\ve_n +\Omega_m | \, t_b \right)
\over  \sinh^{1/2} (x_c / l_{ee} - \pi T t_b )  \sinh^{1/2} (x_a / l_T -
\pi T t_b + x_c / l_{ee} )  },
\label{106}
\end{eqnarray}
the integral over ${\bar t}_b$ again differs only in that $\ve_n +\Omega_m$
changes to $\ve_n$. Using Eqs.~(\ref{105}) and (\ref{106}), we get
\be
{\Delta\sigma_{\rm ME}\over\sigma_{\rm D} }=-A_{\rm ME}\,{l_T\over l}~,
\label{107}
\ee
where $A_{\rm ME}$ is given by the dimensionless integral (check the first
arguments of ${\rm F}$)
\begin{eqnarray}
A_{\rm ME}& = &
\pi \int_{-\infty}^\infty\!{ dz \over \cosh^2  \pi z  }\int_0^\infty\!dx\,
\, \big| {\cal M}(z,x) \big|^2 \,  \big| {\cal M}(z,-x) \big|^2~. \nonumber \\
 \label{108}
\end{eqnarray}
The function ${\cal M} (z,x)$ in Eq.~(\ref{108}) is defined by
\be
{\cal M} (z,x) = {1\over 2 \pi} \,
\int_0^\infty\!\!dy\,{\exp (2 i  z y)  \over
\sinh^{1/2} (y)  \sinh^{1/2} (y +x)}~.
\label{109}
\ee
We have estimated $A_{\rm ME}$ numerically as $A_{\rm ME}\simeq 0.2$.

\end{widetext}

Finally, let us discuss the overall combinatorial factor in
$\Delta\sigma_{\rm ME}$ (which is already included in $A_{\rm ME}$).  Firstly, similarly
to the WL correction, the contribution of diagram (a) in
Fig.~\ref{3diagramsME} should be multiplied by a factor of $2\times2\times2=8$
due to (i) two possible chiralities of the current vertices; (ii) diagrams (b)
and (c) in Fig.~\ref{3diagramsME}; and (iii) two possible anomalous
configurations for each diagram: $x_a\sim l_T \ll x_b$ and $x_b\sim l_T \ll
x_a$.
Secondly, as mentioned above, not only the return after one single
backscattering but rather the entire diffuson ladder contributes to
$\Delta\sigma_{\rm ME}$ [Fig.~\ref{3diagramsME}]. The insertion of the diffuson
into the three-impurity diagram effectively generates an additional
 velocity-vertex correction, which
yields a factor of 1/2 (the ratio of the transport and total scattering times
for the backscattering impurities).

\section{Path-integral method}
\label{VI}
\setcounter{equation}{0}

So far, we have been treating the conductivity correction within the formalism
of Matsubara functional bosonization. This powerful method treats on an equal
footing~\cite{GMP} the real inelastic scattering processes responsible for
dephasing and the virtual transitions responsible for the renormalization
effects. An alternative approach, formulated for the spinless case in
Refs.~\onlinecite{GMPlett} and  \onlinecite{GMP}, consists of two steps. Firstly, disorder is
renormalized by virtual processes with characteristic energy transfer larger
than $T$. What is obtained after the renormalization is an effective
``low-energy'' theory which is free of ultraviolet singularities
characteristic of a LL. The low-energy theory is treated by means of a
path-integral approach, analogous to the one developed in
Ref.~\onlinecite{AAK} for higher-dimensional systems. This method is
particularly convenient for an analysis of inelastic scattering (dephasing) in
problems with a nontrivial infrared behavior. In Ref.~\onlinecite{GMP}, the WL
correction to the conductivity was calculated both within the path-integral
and the functional-bosonization schemes, with identical results. The
path-integral calculation also allows one to ``visualize" the origin of the
dephasing processes in terms of quasiclassical trajectories.

In this section, we present a path-integral analysis of the conductivity
correction in the spinful problem. It turns out that the situation here is
more intricate than in the spinless case, for two reasons: (i) the
characteristic energy transfer in the path-integral calculation is of the
order of $T$ (while it was much smaller than $T$ for spinless electrons); (ii)
because of the SCS, the velocity of the quasiclassical trajectories is not
uniquely defined: the whole interval of velocities between $v$ and $u$
contributes to $\Delta\sigma_{\rm WL}$, see Sec.~\ref{III}. As a result, we will
only be able to reproduce the parametrical dependence of the conductivity
correction but not the numerical prefactor. Nevertheless, this analysis is
useful, since it yields a physically transparent picture of the quantum interference
and dephasing in the spin-charge separated system.

%%%%%%%%%%%%%%%%%%%%%%%%%%%%%%%%%%%%%%%%%%%%%%%%%%%%%%%%
\begin{figure}
\includegraphics[width=8.0cm]{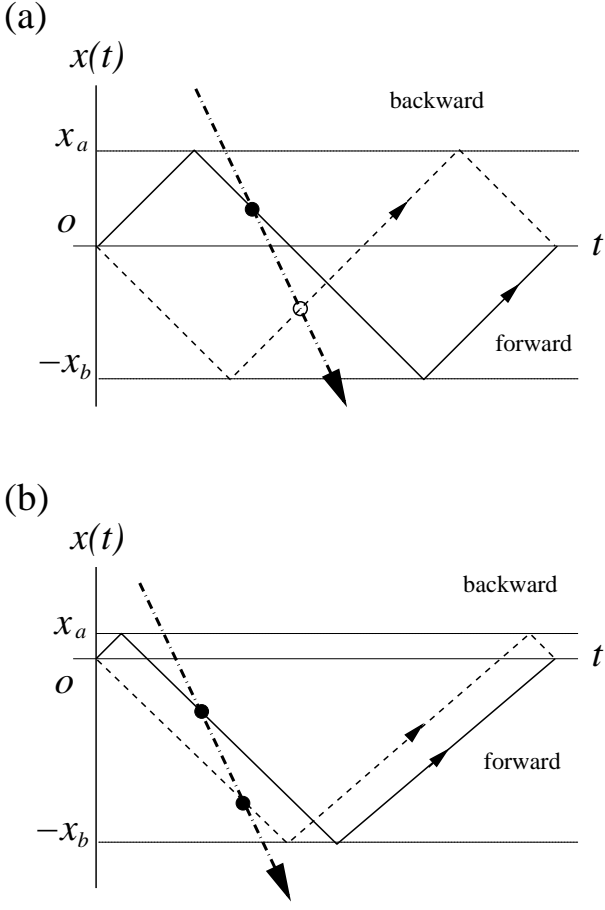}
%\vskip2mm
%\includegraphics[width=8.0cm]{aak11.eps}
\caption{World lines corresponding to the time-reversed trajectories (solid
and dashed lines) with velocity $v^*$ in the three-impurity Cooperon. The dash-dotted line
describes the propagation of a plasmon with velocity $u$.
(a) Regular configuration (Sec.~\ref{IV}) contributing to $\Delta\sigma_{\rm WL}$; (b)
asymmetric configuration (Sec.~\ref{V}) contributing to $\Delta\sigma_{\rm ME}$.
The interaction line gives a contribution to the dephasing action $S$ which is proportional
to $\alpha (N_f-N_b)^2$. Here $N_{f,b}$ is the number of the small-angle intersections
(black dots) of a plasmon line with the forward ($f$) and backward ($b$) paths.
The intersections at a large angle (the plasmon and
electrons moving in opposite directions, unfilled circles) contribute to the
dephasing only at second order in $\alpha$ and hence are neglected.
One sees that $N_f\neq N_b$ for typical configurations (a), whereas for asymmetric configurations
(b) $N_f=N_b$ for most of the plasmon lines. 
For a corresponding plot in the spinless case, see Fig. 7 of Ref.~\onlinecite{GMP}.
}
\label{aak}
\end{figure}
%%%%%%%%%%%%%%%%%%%%%%%%%%%%%%%%%%%%%%%%%%%%%%%%%%%%%%%%%%%%%%%

Since the dominant contribution to the dephasing rate is given by the
$g_4$--processes of scattering between electrons from the same chiral branch,
we will set $g_2=0$ which is consistent with our central approximation
$\alpha_b\to 0$. Furthermore, it is convenient to introduce different
coupling constants,~\cite{sol79,voit94} $g_4^\parallel$ and $g_4^\perp$, for
interaction between electrons with equal and opposite spins, respectively (at
the end, we set $g_4^\parallel=g_4^\perp=g\simeq 2\pi v \alpha$). In view of
the Pauli principle, the $g_4^\parallel$--interaction does not lead to any
real scattering but only renormalizes the velocity of electrons:
\begin{equation}
 v\to v^*=v(1+g_4^\parallel/2\pi v)~.
\label{vstar}
\end{equation}
This allows us to put $g_4^\parallel=0$, simultaneously replacing $v$ by
$v^*$.  All the nontrivial interaction-induced physics is due to
$g_4^\perp$--processes.

Solving the corresponding RPA equations for the interaction propagators,
we get for parallel ($V^{\parallel}$) and anti-parallel ($V^{\perp}$) spins :
\begin{eqnarray}
V_{++}^{\perp} (q, \omega) &=& 2 \pi v^{*} \alpha {(v^{*} q
- \omega)^2 \over (u q - \omega)(v q - \omega )}, \nonumber \\
V_{++}^{\parallel} (q, \omega) &=& - 2 \pi v^{*} \alpha^2 {v^{*} q (v^{*} q
- \omega ) \over (u q - \omega)(v q - \omega )}
\label{RPAVparall}
\end{eqnarray}
and
\be
V^{\perp (\parallel)}_{--} (q, \omega)
= V^{\perp (\parallel)}_{++} (-q, \omega)~,
\ee
where
$
u = v (1+ 2 \alpha)$ and
$v^{*}  =  v (1 + \alpha)$.
We note that $V^{\perp}$ does not enter the path integral, since electron spin
is conserved.

The general expression for the WL dephasing action acquired on the exactly
time-reversed trajectories $x_{\rm b}(t)=x_{\rm f}(t_C-t)$ for the three-impurity  Cooperon
reads~\cite{GMPlett,GMP}
\begin{eqnarray}
\label{Sdeph}
&&S_{ ij}(t_C, \{x(t)\}) \! = \! - T \int_{0}^{t_C} \! d t_1
\int_{0}^{t_C} \! d t_2 \int {d \omega \over 2 \pi} \int {d q \over 2 \pi}
\nonumber \\
&&
\times {{\rm Im} V^{\parallel}_{\mu\nu} (q, \omega) \over \omega }
\exp\left\{  i q  [\, x_i (t_1)-x_j (t_2) ]
- i \omega  [t_1 - t_2] \right\} \nonumber \\
&&
= \! - T \int_{0}^{t_C} \! d t_1 \int_{0}^{t_C} \! d t_2 \,
{\cal F}^{\parallel}_{\mu\nu}[x_i (t_1)-x_j (t_2), \, t_1 - t_2]~,\nonumber \\
&&
\end{eqnarray}
where each of the indices $ i,j$ takes one of the values ``${\rm f}$'' (for the forward
path of the Cooperon) or ``${\rm b}$'' (for the backward path), $\mu={\rm sgn}{\dot
x}_i$ and $\nu={\rm sgn}{\dot x}_j$, and ${\cal
F}^{\parallel}_{\mu\nu}(x,t)$ is the Fourier transform of $\omega^{-1}{\rm Im}
V^{\parallel}_{\mu\nu} (q, \omega)$. The main contribution comes from the
diagonal terms with $i=j$ and $t_1=t_2$, for which $\mu=\nu$, see
Fig.~\ref{aak}a. The imaginary part of the corresponding interaction
propagator is written as
\bea
{\rm Im}\,V_{\pm\pm}^{\parallel} (q, \omega) &=& (\pi v^{*} \alpha)^2 
{\omega\over u v} \nonumber\\
&\times&\left[\ u\, \delta (v q \mp \omega) + 
v\,  \delta (u q \mp \omega)
\right],
\eea
which gives
\bea
{\cal F}^{\parallel}_{\pm\pm} (x,t) = - \pi \,
{(v^{*}\alpha)^2 \over 2 u v} \, \left[\, u\, \delta (x \mp v t)+ v
\, \delta (x \mp u t) \,\right].\nonumber\\
\label{Rxt}
\eea

A graphic illustration of the e-e scattering processes contributing to the dephasing action
$S=2(S_{\rm ff}-S_{\rm fb})$ is presented in Fig.~\ref{aak}. There, we show by the circles and black dots 
the space-time coordinates for which the arguments of the $\delta$-functions in Eq.~(\ref{Rxt}) are zero.
The main contribution to $S$ comes from small-angle intersections (black dots).
For electron trajectories characterized by velocity $v^*$, these give a large factor of the order of $\alpha^{-1}$, 
either $v^*/|u-v^*|$ or $v^*/|v-v^*|$ [cf.\ Eq.~(\ref{tauee-spinfull})]. 
As a result, the action for typical impurity configurations (Fig.~\ref{aak}a) reads
\be
S\sim \alpha^2 T t_C/|\alpha|\sim |\alpha| T t_C.
\ee 

For comparison, in the spinless case\cite{GMPlett,GMP}, intersections 
between the interaction and particle propagators running in the same direction give zero dephasing 
because of the cancellation between the Hartree and exchange terms (Sec.~\ref{IIaa}). On the other hand, 
at large angles the
ballistic (straight line) interaction propagator  always  intersects a pair of (forward and backward)
electron trajectories, which gives zero dephasing as well. 
As a result, only the intersections at large angles that arise from scattering of the 
interaction propagator off disorder contribute to the dephasing rate.\cite{GMPlett,GMP}

\begin{figure}[ht]
\centerline{\includegraphics[width=8.0cm]{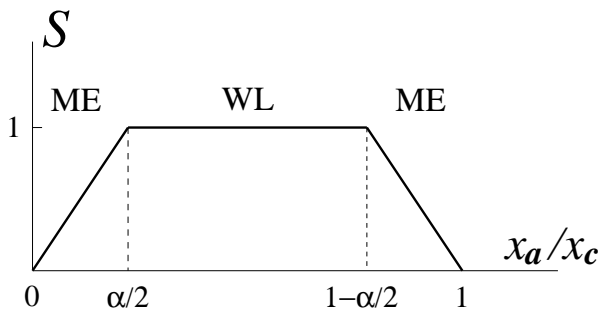}}
\caption{Dephasing action $S$ (in units of $2x_c/l_{ee}$) corresponding to the velocity $v^*$,
Eq.~(\ref{64}), as a function of $x_a$ for fixed $x_c$.
}
\label{action}
\end{figure}

More rigorously, substituting Eq.~(\ref{Rxt}) in Eq.~(\ref{Sdeph}), we find the dephasing
action for the trajectory characterized
by the velocity $v^*$,
\begin{equation}
 S \simeq  \frac{8\pi T}{v}\left\{
\begin{array}{ll}
x_a , & \quad
0 <  x_a < \alpha x_c /2~,
\\[2mm]
\alpha x_c /2, & \quad
\alpha x_c /2
<  x_a <
(1- \alpha/2) \,
x_c~,
\\[2mm]
x_c - x_a, & \quad
(1- \alpha/2) \,
x_c < x_a < x_c~,
\end{array}
\right.
\label{64}
\end{equation}
which is illustrated in Fig.~\ref{action}. Any other velocity between $v$ and
$u$ yields a qualitatively similar action, but with different numerical
factors. This is another indication of the previously discussed fact that all
velocities from this interval contribute to the conductivity correction.
%%%%%%%%%%%%%%%%%%%%%%%%%%%%%%%%%%%%%%%%%%%%%%%%%%%%%%%

One sees from Eq.~(\ref{64}) (middle line) and Fig.~\ref{action} (flat region)
that for typical impurity configurations the dephasing action becomes of the
order of unity at $t_C=2x_c/v^*\sim 1/\alpha T$. This gives a parametric estimate for
the dephasing length $l_\phi \sim \alpha^{-1} l_T$, and for the WL correction
$\Delta\sigma_{\rm WL}/\sigma_{\rm D}\sim -(l_{ee}/l)^2$, in agreement with the results
of Sec.~\ref{IV}. Equation~(\ref{64}) (first and third lines) and Fig.~\ref{action}
also demonstrate
the suppression of the dephasing action in the anomalous (strongly asymmetric)
impurity configurations (Fig.~\ref{aak}b) responsible for the ME,
$\Delta\sigma_{\rm ME}/\sigma_{\rm D}\sim - l_T /l$, as discussed
in Sec.~\ref{V}.

\section{Summary}
\label{VII}
\setcounter{equation}{0}

%%%%%%%%%%%%%%%%%%%%%%%%%%%%%%%%%%%%%%%%%%%%%%%%%%%%%%%%%%%%%%%%%%%

\begin{table*}
\begin{tabular}{|c|c|c|c|c|} \hline \hline
&&&& \\
& ~~ e-e scattering~~ & ~~ dephasing~~ &
WL correction & ~ME correction~ \\
& length $l_{ee}$ & length $l_{\phi}$ &
$\Delta \sigma_{\rm WL} / \sigma_{\rm D} $ & $\Delta \sigma_{\rm ME} / \sigma_{\rm D} $  \\
&&&& \\
\hline
&&&& \\
spinless &&&& \\
%~~electrons~~
& \raisebox{1.5ex}[0pt]{$\displaystyle{{v \over \alpha^2 T}}$} &
\raisebox{1.5ex}[0pt]{$\displaystyle{{1\over\alpha} \left({{v l \over  T}}\right)^{1/2}}$}
& ~~\raisebox{1.5ex}[0pt]{$\displaystyle{  - \Big( {l_{\phi} \over l}
\Big)^2 \ln {l \over l_{\phi}}
\sim  -{ v \ln (\alpha^2 T l/v) \over  \alpha^2 T l } }$}~~
& ~~\raisebox{1.5ex}[0pt]{$\displaystyle{- {l_{ee} \over l} \sim -
{v \over \alpha^2 T l }}$}~~
\\
&&&&  \\
spinful &&&& \\
%electrons
&  \raisebox{1.5ex}[0pt]{$\displaystyle{{v \over \alpha T}}$}  &
\raisebox{1.5ex}[0pt]{$\displaystyle{{v \over \alpha T }}$} &
\raisebox{1.5ex}[0pt]{$\displaystyle{ -\Big( {l_{\phi} \over l}
\Big)^2 \sim - \Big( {v \over \alpha T l } \Big)^2}$}
& \raisebox{1.5ex}[0pt]{$\displaystyle{-{l_T \over l} \sim - {v
\over T l} }$} \\
&&&& \\
\hline \hline
\end{tabular}
\caption{\label{table} Characteristic spatial scales induced by interaction and the
conductivity corrections for spinless and spinful disordered LLs.}
\end{table*}

%%%%%%%%%%%%%%%%%%%%%%%%%%%%%%%%%%%%%%%%%%%%%%%%%%%%%%%%%%%%%%%%%%%

\begin{figure}[h]
\vspace*{3mm}
\centerline{
\includegraphics[width=7.5cm]{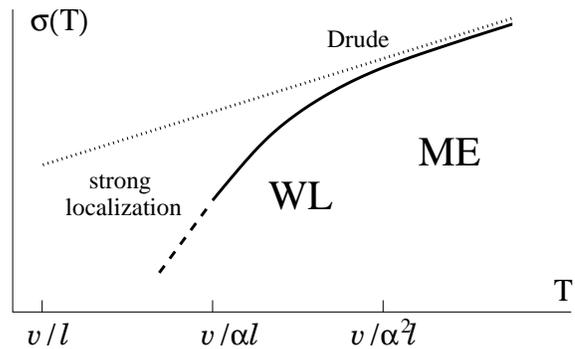}
}
\caption{
Schematic  plot (log-log scale) of the $T$ dependence of the conductivity $\sigma(T)$ of 
spinful electrons.
}
\label{scheme}
\end{figure}

To conclude, we have analyzed, within the functional bosonization formalism,
the quantum interference of interacting electrons in a disordered spinful
LL. Our results are summarized in Table~\ref{table} and in Fig.~\ref{scheme}.

The single-particle properties of fermionic excitations in this model have
been studied in Sec.~\ref{III} in several representations. Two most important
and interrelated features of the single-particle spectral characteristics are:
(i) the SCS, as a result of which the whole range of velocities between the
charge velocity $u$ and the spin velocity $v$ contributes to the spectral
function, and (ii) the single-particle decay on the spatial scale of $l_{ee}$,
Eq.~(\ref{23}). 

In Sec.\ref{IV} we have calculated the leading quantum
interference correction to the conductivity, Eq.~(\ref{60}). The corresponding
dephasing length $l_\phi$ is given by Eq.~(\ref{62}). Two qualitative
differences as compared to the spinless case should be emphasized in this
context.
Firstly, for spinful electrons, the decay length of single-particle excitations
is inversely proportional to the first order in the interaction
strength $\alpha$, $l_{ee}\sim \alpha^{-1} l_T $, whereas $l_{ee} \sim
\alpha^{-2} l_T $ in the absence of spin.  Secondly, the WL dephasing
length $l_\phi$ is equal to the single-particle dephasing length
$l_{ee}$ in the spinful model, whereas $l_\phi$ depends on the
strength of disorder, $l_\phi \sim \sqrt{l_{ee} l}$, for spinless
electrons. 

In Sec.~\ref{V} we have analyzed the contribution to the
Cooperon diagrams of ``nontypical" configurations of disorder with two
impurities being anomalously close to each other. We have shown that
this contribution, Eq.~(\ref{107}), describes the quasiclassical
ME. It is parametrically smaller than the WL term at sufficiently low
$T$ and gives the leading correction to the conductivity in the limit
of high $T$. 

Our findings, demonstrating the strong dependence of the quantum interference
effects on spin in a LL, imply that the Zeeman splitting by magnetic field
should lead to strong effects in the conductivity of a single-channel quantum
wire. Results obtained in this direction will be reported
elsewhere.\cite{magres}

We thank D. Aristov, D. Bagrets, and A. Morpurgo for useful discussions and
gratefully acknowledge support by EUROHORCS/ESF (AGY and IVG), by the DFG
Center for Functional Nanostructures, and by RFBR [Grant No. 06-02-16702] (AGY).


\begin{thebibliography}{100}
\bibitem{remark1} Neglecting $H_{\rm bs}$ is not always justified despite $g_1$
being much smaller than $g_{2,4}$. First, in the presence of spin,
$g_1$-processes ``added" to the clean LL model are a ``marginally relevant" or
``marginally irrelevant" perturbation in the renormalization-group sense,
depending on the sign of the bare $g_1$ (see, e.g.,
Refs.~\onlinecite{sol79,voit94}). For repulsive Coulomb interaction ($g_1>0$),
weak backward e-e scattering is marginally irrelevant, i.e., only leads to a
small renormalization of the effective couplings of the LL theory on
(exponentially) large spatial scales. We neglect the weak renormalization due
to $g_1$-processes altogether. Second, there exist phenomena which are
entirely due to $g_1$-processes. One of these is the spin-related
magnetoresistance of a single-channel quantum wire,\cite{matveev93,magres}
which we will consider elsewhere.\cite{magres} However, for the effects
studied in the present paper, it suffices to deal with forward e-e scattering
only.
\bibitem{footnon} Under the conditions (\ref{103}) and (\ref{103a}),
interaction drops out of the problem only within the approximation
(\ref{45}), corresponding to $g_2=0$. The omitted factors with the
exponent $\alpha_r$ in $Q$ make $Q(0,\tau)$ time-dependent, which
results in the renormalization of impurities---already taken into
account in Eq.~(\ref{44}).
\bibitem{footnote-ladder} This should be contrasted with the regular
configurations, where adding two lines (only diagrams with an odd number of
impurity backscattering lines contribute to the Cooperon) multiplies the
result for the conductivity by a small paramter $(l_{ee}/l)^2$, in view of
dephasing.
\end{thebibliography}
\end{document}